\documentclass[twocolumn]{autart}    %
\usepackage{graphicx}          %

\usepackage{amsmath}
\usepackage{cite}[compress]
\usepackage{dsfont}
\usepackage{amssymb}
\usepackage{siunitx}
\usepackage{caption}
\usepackage{euscript}
\usepackage{xcolor}

\newtheorem{cond}[thm]{Condition}
\newcommand\thefont{\expandafter\string\the\font}
\newcommand{\figref}[1]{\figurename~\ref{#1}}
\newcommand{\tx}[1]{{#1}}
\newcommand{\ar}[2]{\tx{#2 | #1}}

\makeatletter
\g@addto@macro\normalsize{%
  \setlength\abovedisplayskip{7.5pt}
  \setlength\belowdisplayskip{7.5pt}
  \setlength\abovedisplayshortskip{7.5pt}
  \setlength\belowdisplayshortskip{7.5pt}
}

\newcommand{\tripleline}[1]{\begin{tabular}[c]{@{}l@{}l@{}}#1\end{tabular}}

\newcommand{\stox}{{x}}
\newcommand{\stoy}{{y}}
\newcommand{\stou}{{u}}
\newcommand{\stoe}{{e}}
\newcommand{\stoz}{{z}}

\begin{document}

\begin{frontmatter}
\runtitle{Deep Subspace Encoders}  

\title{Deep Subspace Encoders for Nonlinear System Identification} %

\author[Eindhoven]{Gerben I. Beintema$^*$}\ead{g.i.beintema@tue.nl},    %
\author[Eindhoven]{Maarten Schoukens}\ead{m.schoukens@tue.nl},               %
\author[Eindhoven,Budapest]{Roland Toth}\ead{r.toth@tue.nl}  %

\thanks[footnoteinfo]{Implementation of the proposed SUBNET method is available at {\tt https://github.com/GerbenBeintema/deepSI} and the implementation of the simulation study is available at {\tt GerbenBeintema/encoder-automatica-experiments}.}
\thanks[footnoteinfo]{The research was partly funded by the Eötvös Loránd Research Network (grant number: SA-77/2021).}

\address[Eindhoven]{Department of Electrical Engineering, Eindhoven University of Technology, %
Eindhoven, The Netherlands}  %
\address[Budapest]{ Systems and Control Laboratory, Institute for Computer Science and Control, %
Budapest, Hungary.}             %


\begin{abstract}                          %
Using Artificial Neural Networks (ANN) for nonlinear system identification has proven to be a promising approach, but despite of all recent research efforts, many practical and theoretical problems still remain open. Specifically, noise handling and models, issues of consistency and reliable estimation under minimisation of the prediction error are the most severe problems. The latter comes with numerous practical challenges such as explosion of the computational cost in terms of the number of data samples and the occurrence of instabilities during optimization. In this paper, we aim to overcome these issues by proposing a method which uses a truncated prediction loss and a subspace encoder for state estimation. The truncated prediction loss is computed by selecting multiple truncated subsections from the time series and computing the average prediction loss. To obtain a computationally efficient estimation method that minimizes the truncated prediction loss, a subspace encoder represented by an artificial neural network is introduced. This encoder aims to approximate the state reconstructability map of the estimated model to provide an initial state for each truncated subsection given past inputs and outputs. By theoretical analysis, we show that, under mild conditions, the proposed method is locally consistent, increases optimization stability, and achieves increased data efficiency by allowing for overlap between the subsections. Lastly, we provide practical insights and user guidelines employing a numerical example and state-of-the-art benchmark results. \vskip -4mm
\end{abstract} 
\end{frontmatter}

\section{Introduction}\label{sec:intro}

While linear system identification offers both a strongly developed theoretical framework and broadly applicable computational tools, identification of nonlinear systems remains challenging. The wide range of nonlinear behaviours that appear in engineering, reaching from mechatronic systems to chemical and biological systems, poses a challenge in developing generically applicable model structures and identification methods \cite{schoukens2019roadmap}. Hence, numerous nonlinear system identification methods have been proposed over the last decades. Amongst the most popular ones are, linear parameter-varying \cite{Lee1999LPV,toth2010LPV}, Volterra \cite{sliwinski2017Volterra,Birpoutsoukis2017Volterra}, NAR(MA)X \cite{Billings2013NARMAX}, block-oriented \cite{giri2010Block-oriented,Schoukens2017Block-oriented}, and nonlinear state-space \cite{Lennart:deepSSSI,paduart2010polySS,Schon2011NLSS,Schoukens2021SSNN,masti2021autoencoders,beintema2021base-encoder,beintema2020video-encoder,forgione2022SS-state-est} approaches. 

In this paper, we consider the problem of identifying nonlinear systems using \emph{nonlinear state-space} (NL-SS) models since they can represent a broad range of dynamic behaviours and are well applicable for \emph{multiple-input multiple-output} (MIMO) systems~\cite{schoukens2019roadmap}. However, estimation of NL-SS models is rather challenging as the state-variables are often not measurable (hidden Markov model) and the associated optimisation-based training process is prone to local minima and model/gradient instability \cite{DecuyperPNLSSMultipleShooting}. Furthermore, the associated nonlinear state-transition and output functions rapidly grow in complexity with a growing number of states and inputs. If these are parametrized as a linear combination of basis functions, e.g., polynomials as in ~\cite{paduart2010polySS,Decuyper2019PNLSSDecoup}, then this often leads to an explosion of parameters to be able to capture the system dynamics. Also, probabilistic methods such as~\cite{Schon2011NLSS} can become computationally burdensome with increasing numbers of states and inputs or training sequence lengths. Hence, an efficient representation approach for the nonlinearities and a novel estimation concept is required for NL-SS identification.

Deep learning and \emph{artificial neural networks} (ANNs) are uniquely suited to approach the NL-SS identification challenges as they have been shown theoretically and practically to be able to model complex data relations while being computationally scalable to large datasets. Although these benefits inspired the use of state-space neural network models two decades ago~\cite{Suykens1995SSNN}, fully exploiting these properties in NL-SS identification  without major downsides is still an open problem. For instance, careful initialization of the neural network weights and biases partially mitigates the risk of local minima during optimization, but requires additional information, e.g., estimating of a linear approximate model of the system~\cite{Schoukens2021SSNN}. Additionally,~\cite{ribeiro2020smoothness} has shown that multiple shooting smooths the cost function, reducing the number of local minima and improving optimization stability, which has given rise to the use of truncated simulation error cost for ANN based NL-SS estimation \cite{Forgione2021TruncSimulation}. However, the use of multiple shooting approaches comes with the challenge of estimating a potentially large number of unknown initial states for each subsection, resulting in a complexity increase of the  optimisation.  To overcome this problem, auto-encoders have been investigated to jointly estimate the model state and the underlying state-space functions using  one-step-ahead prediction cost \cite{masti2021autoencoders}. However, these approaches fall short of giving accurate long-term predictions due to incorrect noise handling,  they need for tuning sensitive hyperparameters in the composite auto-encoder/prediction-error loss function, and they lack of consistency guarantees.

To overcome these challenges, this paper enhances the subspace encoder-based method for identification of \emph{state-space} (SS) neural networks first introduced in~\cite{beintema2021base-encoder} with an innovation noise model and prove consistency properties. The nonlinear SS model is parametrized with ANNs for flexibility and efficiency in representing the often complex and high-dimensional state-transition and output functions. The model is estimated under a \emph{truncated prediction loss}, evaluated on short subsections. Similarly to multiple shooting, these subsections further improve computational scalability and optimization stability, thereby reducing the importance of parameter initialization. The internal state at the start of each subsection is obtained using a nonlinear \emph{subspace encoder} which approximates the reconstructability map of the SS model and further improves computational scalability and data efficiency. The state-transition and output functions of the SS model and the encoder are simultaneously estimated based on the aforementioned truncated prediction loss function. Finally, \emph{batch optimization} and \emph{early stopping} are employed to further improve the performance of the proposed identification scheme. We demonstrate that the resulting nonlinear state-space identification method is robust w.r.t.~model and gradient instability during training, has a relatively small number of hyperparameters, and obtains state-of-the-art results on benchmark examples.

To summarize, our main contributions are
\begin{itemize}
    \item A novel ANN-based NL-SS identification algorithm that even in the presence of innovation noise disturbances provides reliable and computationally efficient data-driven modelling; 
    \item Efficient use of multiple-shooting based formulation of the prediction loss via co-estimation of an encoder function representing the reconstructability map of the nonlinear model (computational efficiency); 
    \item Proving that the proposed estimator is consistent (statistical validity) and enhances smoothness of the costs function (optimisation efficiency); 
    \item Guidelines for the choice of hyperparameters and a detailed comparison of the proposed method to the state-of-the-art on a widely used identification benchmark.
\end{itemize}

The paper is structured as follows: Section~\ref{sec:preliminaries} introduces the considered data-generating system and identification problem. Section~\ref{sec:method} discusses the proposed subspace encoder method in detail and provides some user guidelines. We theoretically prove multiple key properties of the proposed method in Section~\ref{sec:theory}, and demonstrate state-of-the-art performance of the method on a simulation example and the Wiener--Hammerstein benchmark in Sections~\ref{sec:numerical}-\ref{sec:experimental}, followed by the conclusions in Section~\ref{sec:conclusion}.

\section{Problem setting and preliminaries} \label{sec:preliminaries}

\subsection{Data-generating system} \label{sec:SystemClass}
Consider a discrete-time system with innovation noise that can be represented by the state-space description:
\begin{subequations}\label{eq:SS-inno}
\begin{gather}
    \stox_\tx{k+1} = f(\stox_\tx{k}, \stou_\tx{k}, \stoe_\tx{k}), \label{eq:SS-inno:a} \\
    \stoy_\tx{k} =  h(\stox_\tx{k})  + \stoe_\tx{k}, \label{eq:SS-inno:b}
\end{gather}
\end{subequations}
where $k\in\mathbb{Z}$ is the discrete-time,  $\stoe$ is an i.i.d.~white noise process with finite variance $\Sigma_\mathrm{e}\in \mathbb{R}^{n_\mathrm{y} \times n_\mathrm{y}}$, and $\stou$ is a quasi-stationary input process independent of $\stoe$ and taking values in $\mathbb{R}^{n_\mathrm{u}}$ at each time moment $k$. Additionally, $\stox$ and $\stoy$ are the state and output processes, taking values in $\mathbb{R}^{n_\mathrm{x}}$ and $\mathbb{R}^{n_\mathrm{y}}$ respectively. The functions $f:\mathbb{R}^{n_\mathrm{x}\times n_\mathrm{u} \times n_\mathrm{y}} \rightarrow \mathbb{R}^{n_\mathrm{x}}$ and $h:\mathbb{R}^{n_\mathrm{x}} \rightarrow \mathbb{R}^{n_\mathrm{y}}$, i.e. the state transition and output functions, are considered to be bounded, deterministic maps. Without loss of generality we can assume that $h$ does not contain a direct feedthrough term. By assuming various structures for $f$ and $h$, many well-known noise structures can be obtained such as \emph{nonlinear output noise} (NOE), \emph{nonlinear auto-regressive with exogenous input} (NARX), \emph{nonlinear auto-regressive with moving average exogenous input} (NARMAX) and \emph{nonlinear Box-Jenkins} (NBJ)~\cite{jansson2003ARXsubspace}. For instance, if $f$ does not depend on $\stoe_k$, then a NL-SS model with an OE noise structure is obtained.

For a given sampled excitation sequence $\{ u_k \}_{k=1}^N$ and potentially unknown initial state $x_1\in\mathbb{R}^{n_\mathrm{x}}$, the obtained response of the considered system \eqref{eq:SS-inno} in terms of a sample path realisation is collected into an ordered \emph{input-output} (IO) data set $\mathcal{D}_{N}=\{(u_k,y_k)\}_{k=1}^N$ used for identification. To avoid unnecessary clutter, we will not use different notation for random variables such as $\stoy_k$ defined by \eqref{eq:SS-inno} and their sampled values, but at places where confusion might arise, we will specify which notion is used. 

\subsection{Identification problem}

Based on the given data sequence $\mathcal{D}_{N}$, our objective is to identify the dynamic relation \eqref{eq:SS-inno}, which boils down to the estimation of $f$ and $h$. Note that these functions can not be estimated directly as $x$ and $e$ are not measured. 

To accomplish our objective, notice that $e_{k} = y_{k} - h(x_{k})$ based on \eqref{eq:SS-inno}, hence, by substitution, we get 
\begin{gather}
    x_{k+1} = f(x_{k},u_{k}, y_{k} - h(x_{k}))= \tilde{f}(x_{k},u_{k}, y_{k}).
\end{gather}
Then, for $n\geq 1$, we can write
\begin{subequations}
\label{eq:obsv:01}
\begin{align}
y_k&=h(x_k)+e_k, \\
y_{k+1}&=  (h \circ \tilde f) (x_{k},u_{k}^k, y_{k}^k)+e_{k+1}, \\[-2mm]
&\ \ \vdots \notag \\[-2mm]
y_{k+n} &= (h \circ_n \tilde f) (x_{k},u_{k}^{k+n-1}, y_{k}^{k+n-1}) + e_{k+n},
\end{align}
\end{subequations}
where $\circ$ stands for function concatenation on the state argument, $\circ_n$ means   $n$-times recursive repetition of $\circ$ (e.g., $h \circ_2 \tilde f = h \circ \tilde f \circ \tilde f$ ),  and $u_{k}^{k+n-1}= [\begin{array}{ccc} u_k^\top & \cdots & u_{k+n-1}^\top \end{array}]^\top$ with $y_{k}^{k+n-1}$ similarly defined. More compactly:
\begin{equation} \label{eq:obsv:02}
{y}_{k}^{k+n}=\Gamma_n(x_k,u_{k}^{k+n-1}, y_{k}^{k+n-1})+e_{k}^{k+n}.
\end{equation}
Note that the noise sequence $e_{k}^{k+n}$ is not available in practice, hence, Eq. \eqref{eq:obsv:02} cannot be directly used in estimation. To overcome this problem, we can exploit the i.i.d. white noise assumption on $e_k$ and calculate the expectation of \eqref{eq:obsv:02} w.r.t. $e$ conditioned on the available past data and the initial state $x_k$:  
\begin{multline} \label{eq:state:predictor}
\hat{y}_k^{k+n}=\mathds{E}_e [ {y}_{k}^{k+n} \mid u_{k}^{k+n-1}, y_{k}^{k+n-1}, x_k ] = \\ \Gamma_n(x_k,u_{k}^{k+n-1}, y_{k}^{k+n-1}),
\end{multline}
which is the so called \emph{one-step-ahead} predictor associated with \eqref{eq:SS-inno} and can be computed for the entire sample path realisation in $\mathcal{D}_{N}$, i.e., $\hat{y}_1^{N}=\Gamma_N(x_1,u_{1}^{N-1}, y_{1}^{N-1})$ or, for a specific sample, as $\hat{y}_\tx{n} = \gamma_n (x_{1},u_{1}^{n-1}, y_{1}^{n-1}) $ with $\gamma_n=(h \circ_n \tilde f)$. We can exploit \eqref{eq:state:predictor} to define the estimator by introducing a parametrized form $\Gamma_{N,\theta}$ of the predictor in terms of 
$f_\theta:\mathbb{R}^{n_\mathrm{x}\times n_\mathrm{u} \times n_\mathrm{y}} \rightarrow \mathbb{R}^{n_\mathrm{x}}$ and $h_\theta:\mathbb{R}^{n_\mathrm{x}} \rightarrow \mathbb{R}^{n_\mathrm{y}}$ defined by the parameters $\theta \in \Theta \subseteq \mathbb{R}^{n_\theta}$. 
The classical way to estimate the parameter vector $\theta$ based on a given data set $\mathcal{D}_N$ and ensure that $f_\theta$ and $h_\theta$ accurately represent \eqref{eq:SS-inno} is to minimize the $\ell_2$ loss of the prediction error $\hat{e}_\tx{k} = y_\tx{k} - \hat{y}_\tx{k}$ between the measured samples $y_k$ and the predicted response $\hat{y}_\tx{k}$ by $\Gamma_{N,\theta}$: 
\begin{align}
 \label{eq:simulation-loss}
    V_{\mathcal{D}_\tx{N}}^{\text{pred}}(\theta) &= \frac{1}{N} \sum_{\tx{k=1}}^{N} \left \| y_\tx{k} -  \hat{y}_\tx{k} \right \|^2_2, 
\end{align}
where the initial state $x_1$ is a parameter which is co-estimated with $\theta$.  In case $f_\tx{\theta}$ does not depend on $\hat{\stoe}_\tx{k}$, which corresponds to an OE noise structure, then \eqref{eq:simulation-loss} is equal to the well-known \emph{simulation error} loss function.

The parametrized predictor $\Gamma_{N,\theta}$, can also be written in a state-space form 
\begin{subequations}
\label{eq:model}
\begin{align}
   \hat{\stox}_\tx{k+1} &= f_\tx{\theta}(\hat{\stox}_\tx{k}, \stou_\tx{k}, \hat{\stoe}_\tx{k}),\label{eq:model:state}\\
    \hat{\stoy}_\tx{k} &= h_\tx{\theta}(\hat{\stox}_\tx{k}), %
\end{align}
\end{subequations} 
where $\hat{\stox}$ and $\hat{\stoy}$ are the predicted state and predicted output taking values from $\mathbb{R}^{n_\mathrm{x}}$ and $\mathbb{R}^{n_\mathrm{y}}$ respectively, while $ \hat{\stoe}$ is the prediction error. In fact, \eqref{eq:model} qualifies as the model structure used to estimate \eqref{eq:SS-inno} through the minimization of the identification criterion \eqref{eq:simulation-loss}.     

In the sequel, we will consider $f_\theta$ and $h_\theta$ to be multi-layer \emph{artificial neural networks} (ANNs), parametrized in $\theta$, where each hidden layer is composed from $m$ activation functions $\phi:\mathbb{R} \rightarrow \mathbb{R}$ in the form of $\stoz_{i,j} = \phi(\sum_{l=1}^{m_{i-1}}\theta_{\mathrm{w},i,j,l} \stoz_{i-1,l}+ \theta_{\mathrm{b},i,j})$ where $\stoz_i=\mathrm{col}(z_{i,1},\ldots,z_{i,{m_i}})$  is the latent variable representing the output of layer $1\leq i\leq q$. Here, $\mathrm{col}(\centerdot)$ denotes composition of a column vector. For $f_\tx{\theta}$ with $q$ hidden-layers and linear input and output layers, this means $f_\tx{\theta}(\hat{\stox}_\tx{k}, \stou_\tx{k}, \hat{\stoe}_\tx{k})= \theta_{\mathrm{w},q+1} \stoz_q(k) + \theta_{\mathrm{b},q+1}$     and $\stoz_{0}(k)=\mathrm{col}(\hat{\stox}_\tx{k}, \stou_\tx{k}, \hat{\stoe}_\tx{k})$. The parameters of the state transition and output functions of \eqref{eq:model} are collected in $\theta$. Furthermore, for the remainder of this paper we will assume that $f_\tx{\theta}$ and $h_\tx{\theta}$ are Lipschitz continuous.  Note that assumption is not restrictive for commonly used neural network structures since the activation functions (ReLu, tanh, sigmoid, ...) used for $\phi$ are Lipschitz continuous. Under these considerations, model structure \eqref{eq:model} represents a recurrent neural network and it is also called  \emph{state-space} (SS) ANN in the literature \cite{Suykens1995SSNN,Schoukens2021SSNN}. 

By using the ANNs $f_\tx{\theta}$ and $h_\tx{\theta}$, one can directly compose the feedforward predictor network $\Gamma_{N,\theta}$ and attempt to solve  minimisation of \eqref{eq:simulation-loss} directly. However, this blunt approach can meet with considerable difficulties. In ANN-based identification, minimizing the simulation error, which is a special case of \eqref{eq:simulation-loss} under an OE noise structure, has been observed to result in accurate models \cite{schoukens2019roadmap}, but its major shortcoming is that the computational cost scales at least linearly with $N$. Furthermore, optimization of this cost function is sensitive to local minima and gradient-based methods commonly display unstable behaviour~\cite{ribeiro2020smoothness}. Hence, the problem that we aim to solve in this paper is twofold: \textit{(i)} achieve consistent estimation of \eqref{eq:SS-inno} under innovation noise conditions using the parametrized SS-ANN model \eqref{eq:model} and one-step-ahead prediction \eqref{eq:simulation-loss} and \textit{(ii)} to provide a consistent estimator that drastically reduces the involved computational cost and ensures implementability. 

\section{The subspace encoder method}\label{sec:method}

This section introduces the proposed subspace encoder method that addresses many of the challenges encountered when using classical prediction or simulation error identification approaches for nonlinear state-space models. The proposed approach builds on the introduction of two main ingredients: a truncated prediction loss based cost function and a subspace encoder which is linked to the concept of state reconstructability.

\subsection{Truncated prediction loss}

In order to overcome the computational difficulties in the minimization of \eqref{eq:simulation-loss}, it is an important observation that the main difficulty comes from forward propagation of the state over the entire length of the data set. Hence in the proposed method, which is an extension of our previous work \cite{beintema2021base-encoder}, a truncated form of the $\ell_2$ prediction loss is considered that emulates well the total prediction loss. This truncated form aims to reduce the computational cost by the utilization of parallel computing and to increase optimization stability~\cite{ribeiro2020smoothness}. By selecting subsections of length $T$ (called the truncation length) in the overall time sequence, the prediction loss is calculated on the selected sections: 
\begin{subequations}
\label{eq:encoder}
\begin{align}
    V_{\mathcal{D}_\tx{N}}^{\text{sub}}(\theta) & = \frac{1}{C} \sum_{\tx{t=1}}^{N-T+1} \sum_{\tx{k=0}}^{T-1} \| y_\tx{t + k}-\hat{y}_\ar{t}{t+k}\|^2_2, \\
    \hat{x}_\ar{t}{t+k+1} &= f_\theta(\hat{x}_\ar{t}{t+k},u_\tx{t + k},\hat{e}_\ar{t}{t+k}), \\
    \hat{y}_\ar{t}{t+k} &= h_\theta(\hat{x}_\ar{t}{t+k}),\\
    \hat{e}_\ar{t}{t+k} &= y_\tx{t+k} - \hat{y}_\ar{t}{t+k},
\end{align}
\end{subequations}
where the pipe ($|$) notation is introduced to distinguish between subsections as $(\text{current index} | \text{start index})$, and $C=(N-T+1)T$. If the truncation length is set to $T = N$, then the prediction loss \eqref{eq:simulation-loss} is recovered. 

Formulation \eqref{eq:encoder} addresses both shortcomings of the prediction loss mentioned in Section \ref{sec:intro}. Based on the fact that the predictions can be computed in parallel, only $T$ computations are required to be performed in series hence providing $\mathcal{O}(T)$ computational scaling, which can be considerably smaller than the initial $\mathcal{O}(N)$. Moreover, as is shown in Section \ref{sec:lip}, the use of truncated sections increases the loss function smoothness~\cite{ribeiro2020smoothness}, which both makes gradient-based optimization methods more stable and reduces the effect of parameter initialisation on the optimisation, making the estimation process more reproducible and less varied~\cite{ribeiro2020smoothness}.

The computational cost of the proposed loss function \eqref{eq:encoder} can be further decreased by not summing over all available subsections of the complete data set $\mathcal{D}_N$ for each optimization step, but only over a subset of subsections. This results in a batch formulation of the loss:
\begin{subequations}
\label{eq:enc:loss}
\begin{gather}
    V_{\mathcal{D}_\tx{N}}^{(\text{sub},\text{batch})}(\theta) = \frac{1}{N_{\text{batch}}} \sum_{\tx{t} \in \mathcal{I}} v_\tx{t}, \\
    v_\tx{t} \triangleq \frac{1}{T} \sum_{k=0}^{T-1} \| y_\tx{t + k} - \hat{y}_\ar{t}{t + k}\|^2_2, \label{eq:l-def} \\
    \mathds{\mathcal{I}} \subset \mathbb{I}_{n+1}^{N-T+1}=\{n+1,n+2,...,N\!-\!T\!+\!1\}\, \\ \text{s.t.} \ |\mathcal{I}| = N_{\text{batch}}, \nonumber
\end{gather}
\end{subequations}
which allows for the utilization of powerful batch optimization algorithms such as the Adam optimizer~\cite{kingma2014adam}. Moreover, it is also not necessary to have the complete data set in memory, see \cite{beintema2020video-encoder}, which is a significant advantage in case of large data sets.

An important problem in the minimization of \eqref{eq:encoder} is that there is no expression for the initial state $\hat{x}_\ar{t}{t}$ of each subsection. Considering the initial state of each section to be an optimisation parameter in the minimization of \eqref{eq:encoder} trades new optimisation parameters for the scalability of the cost function (i.e. number of parameters would scale $\mathcal{O}(N)$). This quickly outweighs the benefits of \eqref{eq:encoder}. Hence, to preserve the advantages of the cost function reformulation, an appropriate estimator of the initial state $\hat{x}_\ar{t}{t}$ is required. The next section introduces an encoder-based state estimator based on the concept of state reconstructability.

\subsection{Subspace encoder}
To introduce the proposed encoder, we first require some preliminary notions from nonlinear system theory.
Due to causality of \eqref{eq:SS-inno}, it is a fundamental property of the state that $x_k$ with $k>1$ is completely determined by the past sequence of inputs $\{u_l\}_{l=1}^{k-1}$ and disturbances $\{e_l\}_{l=1}^{k-1}$  together with an initial state $x_1$. In case of state observability of \eqref{eq:SS-inno}, this initial state $x_1$ can be determined based on a future IO sequence~\cite{isidori1985nonlinearcontrol}. The complementary notion of state reconstructability considers the determination of $x_k$ based on a purely past IO sequence~\cite{isidori1985nonlinearcontrol}. The concepts of state observability and reconstructability and the realization theory that builds upon them both for deterministic and stochastic systems form the cornerstones of subspace identification of linear systems and led to many powerful estimation algorithms, see~\cite{katayama2005subspace} for an overview. 

To exploit the concept of observability and reconstructability in deep learning-based identification of \eqref{eq:SS-inno}, consider the result of our derivations in \eqref{eq:obsv:02}. If for an $n\geq 1$, $\Gamma_{n}$ is partially invertible w.r.t $x_k$ on the open sets $\mathbb{X}_0 \subseteq \mathbb{R}^{n_\mathrm{x}}$, $\mathbb{U}_0 \subseteq \mathbb{R}^{n_\mathrm{u}}$, $\mathbb{Y}_0 \subseteq \mathbb{R}^{n_\mathrm{y}}$,  ${\mathbb{E}}_0 \subseteq \mathbb{R}^{n_\mathrm{y}}$, i.e., there exists a $\Phi_n: \mathbb{U}_0^{n} \times \mathbb{Y}_0^{n+1} \times {\mathbb{E}}_0^{n+1}   \rightarrow \mathbb{R}^{n_\mathrm{x}}$ such that $x_k=\Phi_n(u_{k}^{k+n-1}, y_{k}^{k+n}, {e}_{k}^{k+n})$ with $x_k\in\mathbb{X}_0$ and IO signals in these sets, then \eqref{eq:SS-inno} is called \emph{locally observable} on $(\mathbb{X}_0,\mathbb{U}_0,\mathbb{Y}_0, {\mathbb{E}}_0)$ and $\Phi_n$ is called the \emph{observability map} of \eqref{eq:SS-inno} \cite{isidori1995nonlinearcontrolsystems}. Note that if there exists a $(x_*,w_*) \in \mathbb{R}^{n_\mathrm{x}} \times \mathbb{R}^{nn_\mathrm{u} \times (n+1)n_\mathrm{y} \times (n+1)n_\mathrm{y}}$ that $\nabla_{x_*} \Gamma_{n_\mathrm{x}-1}(x_*,w_*)$ is full row rank, then there are open sets $x_\ast \in \mathbb{X}_0$ and $w_\ast \in \mathbb{U}_0^{n} \times \mathbb{Y}_0^{n+1} \times {\mathbb{E}}_0^{n+1}$ such that the partial inverse of $\Gamma_{n}$ exists in terms of an analytic function $\Phi_{n}$ \cite{isidori1995nonlinearcontrolsystems}. Furthermore, if \eqref{eq:SS-inno} is locally observable on $(\mathbb{X}_0,\mathbb{U}_0,\mathbb{Y}_0, {\mathbb{E}}_0)$, then $\Gamma_{n}$ with $n\geq n_\mathrm{x}-1$ has to be partially invertable in the above defined sense. 

Let $\circ_{n}\tilde f$ be a shorthand for $\tilde f$ when $n=1$ and $\tilde f\circ_{n-1}\tilde f$ for $n>1$. Consider 
\begin{subequations}
\begin{align}
x_k&=( \circ_{n}\tilde f) (x_{k-n}, u_{k-n}^{k-1}, y_{k-n}^{k-1}) \\
& =( \circ_{n}\tilde f)  \left(\Phi_n(u_{k-n}^{k-1}, y_{k-n}^{k}, {e}_{k-n}^{k}) , u_{k-n}^{k-1}, y_{k-n}^{k-1}\right) \\
& = \Psi_n( u_{k-n}^{k-1}, y_{k-n}^{k}, {e}_{k-n}^{k}) \label{eq:rec:01}
\end{align}
\end{subequations}
which is called the \emph{reconstructability map}~\cite{isidori1995nonlinearcontrolsystems} of \eqref{eq:SS-inno} as it allows to recover $x_k$ from past measured IO data. Note that the noise sequence $e_{k-n}^{k}$ is not directly available in practice to compute this recovery based on \eqref{eq:rec:01}, but again we can exploit the i.i.d.~white noise property of  $e_k$ to arrive at:
\begin{equation} \label{eq:state:est}
\bar{x}_k=\mathds{E}_e [ x_k \mid u_{k-n}^{k-1}, y_{k-n}^{k} ] = \bar{\Psi}_n(u_{k-n}^{k-1}, y_{k-n}^{k}),
\end{equation}
giving an efficient estimator of $x_k$. In the sequel, we will exploit this concept to formulate an encoder that approximates $\bar{\Psi}_n$.

 As shown in \eqref{eq:state:est}, there exists a state-estimator in the conditional expectation sense for the original system and also the same estimator can be derived for the model structure \eqref{eq:model}. However, the exact calculation of this estimator for a given ANN parametrization of $f_\theta$ and $h_\theta$ is practically infeasible due to the required analytic inversion in terms of $\Phi_n$ and the computation of the conditional expectation of $\Psi_n$ under a given $\Sigma_\mathrm{e}$. Hence, we aim to approximate $\bar{\Psi}_n$ by introducing a nonlinear function $\psi_\eta$ which is co-estimated with $f_\theta$ and $h_\theta$. Since $\psi_\eta$ aims to approximate the subspace reconstructability map \eqref{eq:rec:01} we call it the subspace encoder. Similarly to $f_\theta$ and $h_\theta$ it is also assumed to be Lipschitz continuous: 
\begin{equation} \label{eq:sub:enc}
    \hat{x}_\ar{t}{t} \triangleq \psi_\eta(u_{t-n}^{t-1}, y_{t-n}^{t}).
\end{equation}
Here, $n$ corresponds to the number of past inputs and outputs, i.e. \emph{lag window}, considered to estimate the initial state, while $\eta\in\Lambda\subseteq \mathbb{R}^{n_\eta}$ is the collection of the parameters associated with $\psi_\eta$ in terms of a corresponding ANN with multiple hidden layers. In order to provide an estimator for the initial state of the considered model structure \eqref{eq:model}, the encoder function $\psi_\eta$ is co-estimated with $f_\theta$ and $h_\theta$ by adding the parameters $\eta$ and the estimated initial state using $\psi_\eta$ to the loss function \eqref{eq:encoder}.
\begin{subequations}
\label{eq:encoderloss}
\begin{align}
    V_{\mathcal{D}_\tx{N}}^{\text{enc}}(\theta,\eta) & = \frac{1}{C} \sum_{\tx{t=n+1}}^{N-T+1} \sum_{\tx{k=0}}^{T-1} \| y_\tx{t + k}-\hat{y}_\ar{t}{t+k} \|^2_2, \\
    \hat{x}_\ar{t}{t} &= \psi_\eta(u_{t-n}^{t-1}, y_{t-n}^{t}), \label{eq:encoderloss a} \\
    \hat{x}_\ar{t}{t+k+1} &= f_\theta(\hat{x}_\ar{t}{t+k},u_\tx{t + k},\hat{e}_\ar{t}{t+k}),  \label{eq:encoderloss b} \\
    \hat{y}_\ar{t}{t+k} &= h_\theta(\hat{x}_\ar{t}{t+k}),  \label{eq:encoderloss d} \\
    \hat{e}_\ar{t}{t+k} &= y_\tx{t+k} - \hat{y}_\ar{t}{t+k},  \label{eq:encoderloss e}
\end{align}
\end{subequations}
where now $C=(N-T-n+1)T$ and which again can be formulated as a batch loss function similar to \eqref{eq:enc:loss}.
The used truncated prediction loss and the introduced subspace encoder lead to a model with a deep network structure for estimation, which we call the \emph{subspace-encoder network} (SUBNET). It is graphically summarized in \figref{fig:n-step-encoder-graphic}.

The derivation of the reconstrucability map has shown that based on  $n = n_\mathrm{x}-1$ past inputs and outputs, an effective unbiased estimator of the initial state $x_{t|t}$ can be achieved. While $n = n_\mathrm{x}-1$ is often the minimal required number of past IO samples to obtain an unbiased estimator, the variance of estimate $\hat x_{t|t}$  can be rather significant and further reduced by increasing $n$. The underlying mechanism is similar to the concept of minimum variance observers~\cite{darouach1997minimumvariance} that provide statically efficient state estimation using $n > n_\mathrm{x}-1$ input and outputs lags. Besides of showing empirically this effect in Section \ref{sec:experimental}, a deeper theoretical exploration of the variance optimal choice of $n$ is not within the scope of this paper. 

\subsection{Parameter estimation}

\begin{figure}
    \centering
    \includegraphics[width=0.95\linewidth]{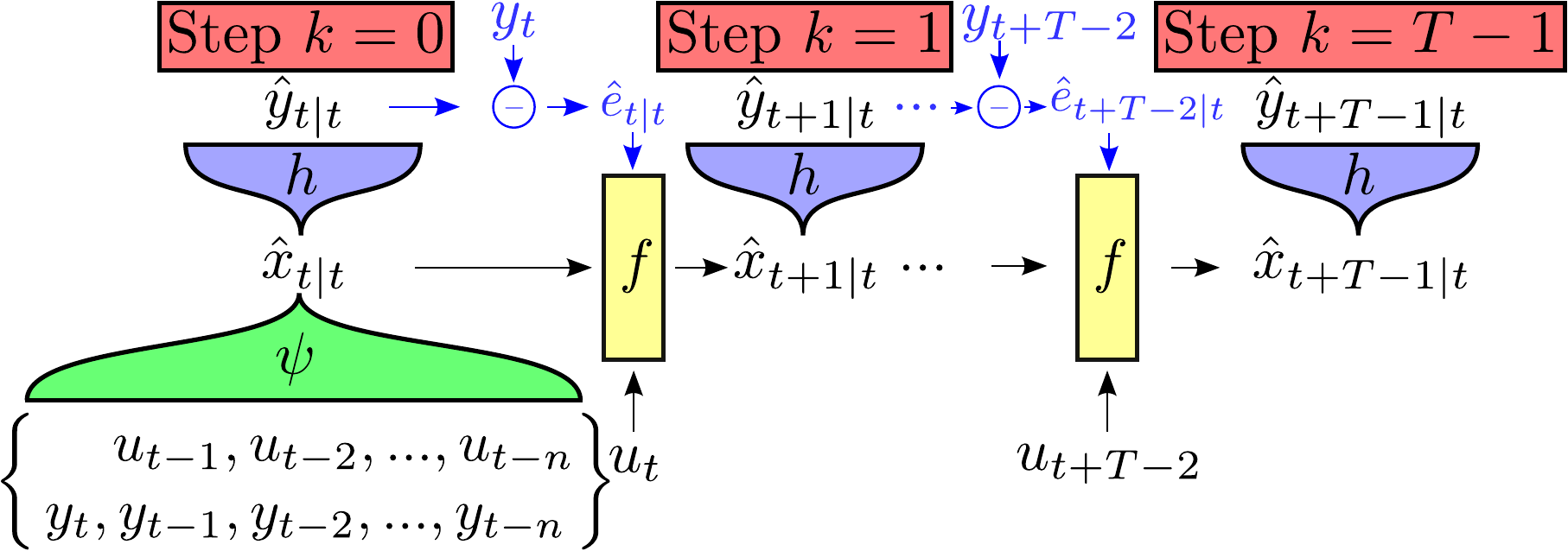}
    \caption{Overall SUBNET structure: the subspace encoder $\psi_\eta$ estimates the initial state at time index $t$ based on past inputs and outputs, then the state is propagated through  $f_\theta$ and $h_\theta$ multiple times until the truncation length $T$. The parts marked in blue constitute the innovation noise process.}
    \label{fig:n-step-encoder-graphic}
\end{figure}

To obtain a model estimate in terms of the SUBNET structure through the minimization of the loss \eqref{eq:encoderloss}, the following steps are executed: \textit{(i)} random initialization of all networks in Fig.~\ref{fig:n-step-encoder-graphic} by an efficient approach such as the Xavier method~\cite{glorot2010Xavierinit}, \textit{(ii)} for the given normalized data (or batches of data) the loss is computed while the computation graph with intermediate values are saved in memory (this uses $\mathcal{O}(N_\text{batch} T)$ memory), \textit{(iii)}  the gradient of the loss is computed by back-propagation using the computation graph obtained in Step \textit{(ii)}, \textit{(iv)} the network parameters are updated by a stochastic gradient optimization method like Adam~\cite{kingma2014adam}, \textit{(v)}  iteration is continued till convergence or cross-validation based early stopping.

\subsection{User guidelines}

The subspace encoder has a number of hyper-parameters that need to be chosen based on the to-be-identified system at hand. Hence, a few guidelines are provided based on insights obtained from theoretical analysis, numerical analysis and practical experience. 

\begin{itemize}
    \item Choose $T$ to be a few times the largest characteristic time scale for stable data-generating systems. For such a choice, the truncated prediction loss \eqref{eq:encoderloss} provides a close approximation of the 'regular' prediction error at a low computational cost. 
    \item $n_\mathrm{x}$ and $n$ need to be chosen as at least the effective order and 
    lag (minimal reconstruction order) of the system, respectively. Furthermore, one can choose different lags $n_a$ for past $y$ and $n_b$ for past $u$ and increase $n$ to reduce the variance of the initial state estimate. 
    \item The choice of the ANN architectures (number of layers $q$, activation functions per layer $m$, type of activation functions) used to parametrize $f_\theta$, $g_\theta$ and $\psi_\eta$ are system dependent. However, overfitting on the data caused by the choice of an over-parametrized architecture is suppressed by the used innovation noise model structure, regularisation induced by the overlapping subsections, early stopping and batch optimisation. Hence, a suggested baseline is to use 2 hidden layer networks with 64 nodes per layer, tanh activation and a linear bypass (similar to a residual component).
    \item IO normalization is essential to make the signals involved  in the estimation to be zero-mean and have standard deviation of one. This is required as IO normalization is a key assumption in many parameter initialization methods (e.g. Xavier initialization~\cite{glorot2010Xavierinit}) and the ``active'' range of many activation functions is also close to a range around zero with a width of 1. After estimation, to remove normalisation, a back scaling is added to the resulting model estimate.
    \item The batch size should be the smallest size which only marginally compromises the data throughput speed (i.e. training samples processed per second) and further reduced to increase regularization effects of batch-optimization. This guideline is according to the current consensus in the ML community. The baseline is $256$, but this is data and architecture dependent.  
    \item For all our experiments a fixed learning rate of $10^{-3}$ using the Adam optimizer has been sufficient. The model quality can be further improved by using early stopping and returning the model of the epoch which had the lowest validation error. 
\end{itemize}

\subsection{Comparison to the state-of-the-art}
Contrary to other approaches that use an encoder function such as~\cite{masti2021autoencoders}, which is based on a modified auto-encoder to learn the latent state and a 1-step ahead prediction loss to learn the system dynamics, we do not need to introduce any additional loss function elements to fit the encoder function. Intuitively, a more accurate estimate of the state automatically reduces the transient error and hence the mismatch between the measurements and the predicted model output. Thus, reducing the transient also reduces the truncated prediction loss, which makes it superfluous to introduce any additional cost function terms.  

The proposed estimation method can be also related to multiple shooting methods~\cite{bock1981multipleshooting}. Multiple shooting also sub-divides the time series into multiple sections and adds the initial state of each section to the parameter vector together with additional constraints~\cite{decuyper2020multipleshootingwithpar}. Compared to this method, our proposed method uses the subspace encoder to directly estimate the initial state from past inputs and outputs for each section. As a consequence, the computational complexity does not increase for an increasing number of sections. Furthermore, our formulation uses overlapping subsections whereas multiple shooting does not make use of overlap. In Section \ref{sec:data-efficiency}, we prove that section overlap increases data efficiency.

Truncated back-propagation through time (truncated BPTT)~\cite{tallec2017TBPTT} also sub-divides the time series, but by truncating the gradient calculation at a truncation length to stabilize the gradient. This still requires a full pass over the time series data which can be computationally expensive and still unstable (value explosion) for large data sets. Furthermore, it adds extra bias and/or variability to the gradient estimate, which is not the case with the proposed subspace encoder method. 

Lastly, the subspace encoder function not only qualifies as a reconstructability map, but also as a state observer. Hence, the encoder can be used to kick-start simulations on possibly unseen data sets. In particular, \emph{nonlinear model predictive control} (NMPC) relies on accurate few-step-ahead prediction models and state estimates \cite{allgower2012nonlinearMPC}, which makes the combined SUBNET structure with the encoder based observer readily applicable for MPC. 

\section{Theoretical analysis}
\label{sec:theory}

In this section, we show key theoretical properties of the proposed encoder method in terms of consistency corresponding to statistical validity of the estimator, loss function smoothness that implies optimisation efficiency, and data efficiency resulting from allowed overlaps in the subsections. 

\subsection{Consistency of the estimator}

The notion of consistency, as defined in \cite{ljung1978convergence}, expresses that the resulting model estimates tend to an equivalent representation of the system that generated the data when the number of data points tends to infinity. In other words, the model estimate is asymptotically unbiased and  converges asymptotically to a true model of the system. Thus in this section, we will show consistency of the SUBNET approach relying on the results of \cite{ljung1978convergence}. 

\textit{Data-generating system:} To show consistency, we need to introduce some conditions on the data-generating system.
As we discussed, the true  system \eqref{eq:SS-inno} can be reformulated in a 1-step-ahead predictor form given by \eqref{eq:state:predictor}. For $k\geq 1$, let $\EuScript{W}_{[1,k]}$ denote the $\sigma$-algebra generated by the random variables $(u^k_1,e^k_1)$ and let $\mu_\mathrm{w}:\EuScript{W}_{[1,k]}\rightarrow [0,1]$ denote the associated probability measure. Furthermore, as $f$ and $h$ are deterministic, define 
\begin{multline}
\hspace{-2mm}\mathfrak{B}\! =\!\left\{(y_1^\infty,x_1^\infty,u_1^\infty,e_1^\infty) \in (\mathbb{R}^{n_\mathrm{w}})^\mathbb{N} \mid   (u_1^\infty,e_1^\infty)\in \EuScript{W}_{[1,\infty]},  \right. \\[0.5mm] \left.  \text{ and } (y_k,x_k,u_k,e_k) \text{ satisfies \eqref{eq:SS-inno} } \forall k\in\mathbb{N}
\right\},
\end{multline}
with $n_\mathrm{w}=n_\mathrm{y} + n_\mathrm{x} + n_\mathrm{u} + n_\mathrm{y}$, being the sample path behavior, i.e., the set of all solution trajectories, of \eqref{eq:SS-inno}. Note that by defining the $\sigma$-algebra $\EuScript{B}$ over $\mathfrak{B}$ and an appropriate probability measure $\mu_\mathrm{b}$, the stochastic behavior of \eqref{eq:SS-inno} can be fully represented, see \cite{jw13}.  

Let $\mathfrak{B}_{[k_\mathrm{o},k]}$ and $\EuScript{B}_{[k_\mathrm{o},k]}$ be the restriction of $\mathfrak{B}$ and $\EuScript{B}$ to the time interval $[k_\mathrm{o},k] \subseteq \mathbb{N}$ with $k\geq k_\mathrm{o}$, respectively. Then, for a given sample path $\{(y_k,x_k,u_k,e_k)\}_{k=k_\mathrm{o}}^\infty \in \mathfrak{B}_{[k_\mathrm{o},\infty]}$ of \eqref{eq:SS-inno} with  $x(k_\mathrm{o})=x_\mathrm{o}$,  $\{(\tilde y_k, \tilde x_k,u_k,e_k)\}_{k=k_\mathrm{o}}^\infty \in \mathfrak{B}_{[k_\mathrm{o},\infty]}$ corresponds to the response of \eqref{eq:SS-inno} for the perturbed state value $\tilde x(k_\mathrm{o})=\tilde x_\mathrm{o}$ at time moment  $k_\mathrm{o}\in\mathbb{N}$ subject to the same input and disturbance as the nominal state response. Based on these, the following stability condition is formulated:

\begin{cond}[Incremental exp. output stability]
\label{assum:S1}
    The data-generating system \eqref{eq:obsv:02} is  (globally) incrementally exponentially output stable, meaning that for any $\delta> 0$, there exist a $0\leq C(\delta)<\infty$ and a $0\leq\lambda < 1$ such that
    \begin{align}
        \mathds{E}_e [ \|y_k - \tilde{y}_k\|_2^4 ] < C(\delta)  \lambda^{k-k_\mathrm{o}}, \quad \forall k\geq k_\mathrm{o}
    \end{align}
    under any $k_\mathrm{o}\geq1$,  $ x_\mathrm{o},\tilde{x}_\mathrm{o}\in\mathbb{R}^{n_\mathrm{x}}$ with $\|x_\mathrm{o}-\tilde{x}_\mathrm{o}\|_2<\delta$ and $(u_1^\infty,e_1^\infty)\in \EuScript{W}_{[1,\infty]}$, where the random variables $y_k$ and $\tilde{y}_k$ belong to $\EuScript{B}_{[k_\mathrm{o},\infty]}$ with the same $(u_k,e_k)$, but with $x_{k_\mathrm{o}}=x_\mathrm{o}$ and $\tilde x_{k_\mathrm{o}}=\tilde x_\mathrm{o}$.
\end{cond}

\textit{Model Set:} 
The considered SUBNET model \eqref{eq:encoderloss a}-\eqref{eq:encoderloss e} corresponds to a model structure $M_\xi$ parameterized by a finite-dimensional parameter vector $\xi = [\begin{array}{cc} \theta^\top & \eta^\top \end{array}]^\top$ that is restricted to vary in a compact set $\Xi\subset \mathbb{R}^{n_\xi}$. The resulting model set is $\mathcal{M}=\{ M_\xi \mid \xi\in\Xi\} $. 
For each $\xi \in \Xi$, the SUBNET model $M_\xi$ with a given encoder lag $n\geq 1$, can be written in a 1-step-ahead predictor form 
\begin{equation} \label{eq:subnet:pred}
    \hat{y}_\ar{t}{t+k} = \hat{\gamma}_{k}(\xi,y^{t+k-1}_{t-n},u^{t+k-1}_{t-n}).
\end{equation}
For $\mathcal{M}$, two important conditions are considered.

\begin{cond}[Differentiability] 
\label{assum:differentiability} The 1-step-ahead predictor
$\hat{\gamma}_{k}: \mathbb{R}^{n_\theta+n_\eta}\times \mathbb{R}^{(n_\mathrm{y}+n_\mathrm{u})(n+k)}\rightarrow \mathbb{R}^{n_\mathrm{y}}$ is differentiable with respect to $\xi$ for all $\xi\in\breve\Xi$, where $\breve{\Xi}$ is an open neighborhood of $\Xi$. \end{cond}

Next, we require the influence of delayed inputs and outputs on the predictor map $\hat{\gamma}_k$ to be exponentially decaying with a number of delays to assure the convergence of the predictor. This is formalized as follows;

\begin{cond}[Predictor convergence]
\label{assum:M1}
\newcommand{\ov}[2]{\overset{#2}{#1}} There exist a $0\leq C<\infty$ and a $0\leq \lambda<1$ such that, 
for any $k\geq0$ and $\xi\in\breve\Xi$, where $\Xi $ is an open neighborhood of $\Xi$, the deterministic predictor map $\hat{\gamma}_k$ under Condition \ref{assum:differentiability} satisfies %
    \begin{multline} \label{eq:M1}
        \|\hat{\gamma}_{k}(\xi,{u}_{-n}^{k-1},{y}_{-n}^{k-1})-\hat{\gamma}_{k}(\xi,\tilde{u}_{-n}^{k-1},\tilde{y}_{-n}^{k-1})\|_2  \\
        \leq C \sum_{s=-n}^{k} \lambda^{k-s} \left(\|{u}_s-\tilde{u}_s\|_2 + \|{y}_s-\tilde{y}_s\|_2 \right) , 
\end{multline}
for any $(u_{-n}^{k-1},y_{-n}^{k-1}),(\tilde u_{-n}^{k-1},\tilde y_{-n}^{k-1})\in\mathbb{R}^{(n_\mathrm{y}+n_\mathrm{u})(n+k)}$ and
    \begin{align}
        \|\hat{\gamma}_{k}(\xi,0_{-n}^{k-1},0_{-n}^{k-1})\|_2 \leq C,
    \end{align}
where $0_{t-n}^{t+k-1} = [\ 0\ \cdots\ 0\ ]^\top$. Furthermore, \eqref{eq:M1} is also satisfied by $\frac{\partial}{\partial \phi}\hat{\gamma}_k(\xi,y^{k-1}_{-n},u^{+k-1}_{-n})$. 
\end{cond}

\textit{Convergence:} Under the previous considerations, convergence of the SUBNET estimator can be shown, which is a required property to show consistency.  

\begin{thm}[Convergence] \label{lem:convergence}
Consider system \eqref{eq:SS-inno} satisfying Condition \ref{assum:S1} with a quasi-stationary $u$ independent of the white noise process $\stoe$. Let the set of models $\mathcal{M}$ defined by the model structure \eqref{eq:encoderloss a}-\eqref{eq:encoderloss e} for $\forall\xi\in\Xi$ satisfy Conditions \ref{assum:differentiability} and \ref{assum:M1}. Then
    \begin{align} \label{eq:conv}
        \mathrm{sup}_{(\theta, \eta)\in \Xi} \left \|V_{\mathcal{D}_\tx{N}}^{\mathrm{enc}}(\theta, \eta) - \mathds{E}_e [ V_{\mathcal{D}_\tx{N}}^{\mathrm{enc}}(\theta, \eta)]\right\|_2 \rightarrow 0,
    \end{align}
    with probability 1 as $T,N \rightarrow \infty$ and the sequence of functions $\mathds{E}_e [V_{\mathcal{D}_\tx{N}}^{\mathrm{enc}}(\theta, \eta)]$ is equicontinuous in $\xi
    \in \Xi$.
\end{thm}
\begin{pf} 
The mean squared prediction error identification criterion used in \eqref{eq:encoderloss} satisfies  Condition C1 in \cite{ljung1978convergence}, hence the proof of \cite[Lemma 3.1]{ljung1978convergence} applies for the considered case.  \hfill $\blacksquare$ \vspace{-3mm}
\end{pf}

\textit{Consistency:} In order to show consistency, we need to assume that the system is part of the model set. Consider the  state-reconstructability map ${\Psi}_n$ in \eqref{eq:rec:01} for the data-generating system \eqref{eq:SS-inno} with $n\geq n_\mathrm{x}$. Note that
\begin{align}
y_\tx{t+k} &= \gamma_k ({\Psi}_n(u_{t-n}^{t-1}, y_{t-n}^{t},e_{t-n}^{t}),u_{t}^{t+k-1}, y_{t}^{t+k-1})+e_{t+k},\notag \\
& = \breve{\gamma}_k(u_{t-n}^{t+k-1}, y_{t-n}^{t+k-1}, e_{t-n}^{t})+e_{t+n} 
\end{align}
for any $t,k\geq 0$, where $\gamma_k$ is according to \eqref{eq:state:predictor}, i.e.,  $\gamma_k=(h \circ_k \tilde f)$. Then,
\begin{align}
\bar{y}_\ar{t}{t+k} &=\mathds{E}_e[\breve{\gamma}_k(u_{t-n}^{t+k-1}, y_{t-n}^{t+k-1}, e_{t-n}^{t})] \notag \\
& = \bar{\gamma}_k (u_{t-n}^{t+k-1}, y_{t-n}^{t+k-1}), \label{eq:pred:opt}
\end{align}
is the optimal one-step-ahead predictor associated with \eqref{eq:SS-inno} under an $n$-lag based reconstructability map.

\begin{defn}[Equivalence set] \label{defn:class} For a  given model structure $M_\xi$ with encoder lag $n\geq n_\mathrm{x}$, predictor $\hat{\gamma}_{k}(\xi,\centerdot)$ (see \eqref{eq:subnet:pred}) and $\xi\in\Xi \subset \mathbb{R}^{n_\xi}$, the set of equivalent models with the data-generating system \eqref{eq:SS-inno} in the one-step-ahead prediction sense \eqref{eq:pred:opt} is defined as 
    \begin{equation}
        \Xi_{\ast} = \left \{ \xi \in \Xi \mid \hat{\gamma}_k(\xi,\centerdot)=\bar\gamma_k(\centerdot),\ \forall k \geq 0   \right \}. \label{eq:xi-ast-def}
    \end{equation}
\end{defn}
Note that if $\Xi_{\ast} \neq \varnothing$, then there exists a $M_{\xi_\ast}\in\mathcal{M}$ that is equivalent with \eqref{eq:SS-inno}. In this case, we call the considered model set to be sufficiently rich to contain an equivalent realisation of the data-generating system. Next we need to ensure that under the given observed data from \eqref{eq:SS-inno}, we can distinguish non-equivalent models in $\mathcal{M}$.  

\begin{cond}[Persistence of excitation]
    \label{assum:PE} Given the model set $\mathcal{M}=\{ M_\xi \mid \xi\in\Xi\} $ with $n\in\mathbb{N}$ and the associated  
     $V_{\mathcal{D}_\tx{N}}^{\mathrm{enc}}$
    in terms of \eqref{eq:encoderloss} with $0\leq T\leq N$, we call the input sequence $u_1^N$ in $\mathcal{D}_\tx{N}$ generated by \eqref{eq:SS-inno} to be  weakly persistently exciting, if for all pairs of parameterizations given by $\left( \theta_1, \eta_1 \right) \in \Xi$ and $\left( \theta_2, \eta_2\right)\in \Xi$ for which the function mapping is unequal, i.e., $V_{(\cdot)}^{\mathrm{enc}}(\theta_1,\eta_1) \neq V_{(\cdot)}^{\mathrm{enc}}(\theta_2,\eta_2)$, we have
    \begin{align}
        V_{\mathcal{D}_\tx{N}}^{\mathrm{enc}}(\theta_1,\eta_1) \neq V_{\mathcal{D}_\tx{N}}^{\mathrm{enc}}(\theta_2,\eta_2).
    \end{align}
    with probability 1.
\end{cond}

To show consistency, we also require that any element of $\Xi_{\ast}$ has minimal cost.

\begin{property}[Minimal cost] \label{thm:minCost} If $\mathcal{M}$ is sufficiently rich, then for any $\xi = [\ \theta^\top\ \eta^\top\ ]^\top \in\Xi$ and $\phi_\ast = [\ \theta_\ast^\top \ \eta_\ast^\top\ ]^\top \in \Xi_\ast$ the encoder loss in \eqref{eq:encoderloss} has the following property: 
\begin{gather} \label{eq:mincost}
    \lim_{T,N \rightarrow \infty} V_{\mathcal{D}_\tx{N}}^{\mathrm{enc}}(\theta_\ast, \eta_\ast) \leq \lim_{T,N \rightarrow \infty} V_{\mathcal{D}_\tx{N}}^{\mathrm{enc}}(\theta, \eta) 
\end{gather}
with probability 1. 
\end{property}
\begin{pf}
Since $\mathds{E}_e [ V_{\mathcal{D}_\tx{N}}^{\mathrm{enc}}(\theta_\ast, \eta_\ast) ]$ exists as shown in Theorem \ref{lem:convergence}, it is sufficient to show that 
\begin{multline}
\lim_{T\rightarrow\infty} \frac{1}{T} \sum_{k=0}^{T-1} \|\hat{y}^\ast_\ar{t}{t+k} - y_{t+k} \|_2^2 \leq \\[-2mm] 
\lim_{T\rightarrow\infty} \frac{1}{T} \sum_{k=0}^{T-1} \|\hat{y}_\ar{t}{t+k} - y_{t+k} \|_2^2    
\end{multline}
for all $t$ where $\hat{y}^\ast_\ar{t}{t+k}= \hat{\gamma}_{k}(\xi_\ast,y^{t+k-1}_{t-n},u^{t+k-1}_{t-n})$. By the law of large numbers, as $T\rightarrow \infty$, the sample distribution of $\{e_{t+k}\}_{k=0}^{T-1}$ will converge to the original white noise distribution of $e$ with finite variance $\Sigma_e$ and with probably 1. Thus, it is sufficient to show that 
\begin{equation}
\mathds{E}_e [ \|\hat{y}^\ast_\ar{t}{t+k} - y_{t+k} \|_2^2 ] \leq \\ 
\mathds{E}_e [ \|\hat{y}_\ar{t}{t+k} - y_{t+k} \|_2^2 ]    
\end{equation}

which can be expanded with $y_{t+k} = h(x_{t+k}) + e_{t+k}$ as
\begin{multline}
\mathds{E}_e [ \|\hat{y}^\ast_\ar{t}{t+k} - y_{t+k} \|_2^2 ]  = \mathds{E}_e [ \|\hat{y}^\ast_\ar{t}{t+k} - h(x_{t+k})\|_2^2 ] - \\
\mathds{E}_e [ 2 (\hat{y}^\ast_\ar{t}{t+k} - h(x_{t+k}))\cdot e_{t+k} ] + \mathds{E}_e [ \|e_{t+k} \|_2^2 ].        
\end{multline}
The second term of this expansion equals to zero since $e_{t+k}$ is uncorrelated to $(\hat{y}^\ast_\ar{t}{t+k} - h(x_{t+k}))$ and $e_{t+k}$ is  zero-mean. Furthermore, the first term is also zero since in terms of Definition \ref{defn:class}, $\hat{y}^\ast_\ar{t}{t+k}= \hat{\gamma}_{k}(\xi_\ast,y^{t+k-1}_{t-n},u^{t+k-1}_{t-n})$  is equal to $\bar{y}_\ar{t}{t+k}$ in \eqref{eq:pred:opt}.
Hence, 
\begin{equation}
    \mathds{E}_e [ \|\hat{y}^\ast_\ar{t}{t+k} - y_{t+k} \|_2^2 ] = \| \Sigma_e \|_2^2
\end{equation}
which is irreducible and thus minimal. \hfill $\blacksquare$
\end{pf}

\begin{thm}[Consistency] \label{lem:consistency}
Under the conditions of Theorem \ref{lem:convergence}, Condition \ref{assum:PE} and Property \ref{thm:minCost}, 
    \begin{align}
        \underset{T,N \rightarrow \infty}{\lim} \hat{\xi}_N \in \Xi_{\ast}
    \end{align}
    with probability 1, where
    \begin{align}
        \hat{\xi}_N = \arg \min_{\xi \in \Xi} %
        V_{\mathcal{D}_\tx{N}}^{\mathrm{enc}}(\theta, \eta).
    \end{align}
\end{thm}
\begin{pf} 
See Lemma 4.1 in \cite{ljung1978convergence}. Note that the squared loss function \eqref{eq:encoderloss} fulfils Condition (4.4) in \cite{ljung1978convergence}. \hfill $\blacksquare$ \end{pf}

\subsection{Increased cost smoothness due to truncation}
\label{sec:lip}

Next, we show that the considered estimation structure and the truncated prediction loss increase the smoothness of the cost function, which potentially makes the optimization process for model estimation more stable and less prone to get stuck in local minima~\cite{ribeiro2020smoothness}. For this purpose, we investigate the smoothness of the encoder loss function by the means of the Lipschitz-continuity analysis. The Lipschitz constant $L_{\text{enc},T}\geq 0$ for the considered loss function is defined as 
\begin{multline}
    \| V^{\text{enc},T}_{\mathcal{D}_\tx{N}} (\theta_1, \eta_1) - V^{\text{enc},T}_{\mathcal{D}_\tx{N}} (\theta_2, \eta_2)\|_2^2 \\ 
    \leq  L_{\text{enc}, T}^2 (\| \theta_1 - \theta_2 \|_2^2 + \| \eta_1 -  \eta_2 \|_2^2) 
\end{multline}
with $[\ \theta^\top \ \eta^\top\ ]^\top \in\Xi = \Xi_\theta \times \Xi_\eta$, assuming that $\Xi$ is not only compact, but it is also convex. Here, the $T$ dependence of the loss function is added explicitly. Since $L_{\text{enc}, T}$ bounds the slope of the function, it provides insight into the smoothness of the cost function as $T$ changes. By the following theorem, we show that smoothness of $V^{\text{enc},T}_{\mathcal{D}_\tx{N}}$ can decrease exponentially with increasing $T$.

\begin{thm}
Assume that $f_\tx{\theta}$, $h_\tx{\theta}$ and $\psi_\eta$ are Lipschitz continuous with Lipschitz constants $L_f$, $L_h$ and $L_\psi$. Then, $L_{\text{enc}, T}$ and $L_{\text{enc}, T}'$, representing the Lipschitz constant of the derivative of $V^{\text{enc},T}_{\mathcal{D}_\tx{N}}$, scale as 
\begin{gather}
    L_{\text{enc}, T} = \mathcal{O}(L_{\Tilde{f}}^{2 T});\ \ L_{\text{enc}, T}' = \mathcal{O}(L_{\Tilde{f}}^{3 T}). \label{eq:lip-scaling-encoder}
\end{gather} 
if $ L_{\Tilde{f}} = L_f \sqrt{1+ L_h^2} > 1$. 
\end{thm}
\begin{pf}
For Lipschitz-continuous functions $q(x)$ and $p(x)$, two known properties of the Lipschitz constant are: \textit{(i)} the \textit{sum} of two functions $q(x) + p(x)$ has the Lipschitz constant $L_q + L_p$ and \textit{(ii)} the \textit{multiplication} of two functions  $q(x) p(x)$ has the Lipschitz constant $L_q m_p + L_p m_q$, where $m_q$ is the maximum of $q(x)$ on the considered compact set $x \in \Xi_x$ and $m_q$ is similarly defined. 

The Lipschitz constants of $f_\tx{\theta}$, $h_\tx{\theta}$ and $\psi_\eta$ are defined by the following relations:
\begin{subequations}
\begin{equation}
 \| h_{\theta_1}(x) - h_{\theta_2} (\tilde x) \|_2^2 \leq L_h^2 (\| \theta_1 - \theta_2 \|_2^2 + \| x - \tilde{x} \|_2^2), \label{lip:1}\vspace{-2mm}
\end{equation}
\begin{multline}
    \| f_{\theta_1}(x,u,y-h_{\theta_1}(x)) - f_{\theta_2}(\tilde{x},u,y-h_{\theta_2}(\tilde{x})) \|_2^2 \\
    \leq L_f^2 (\| \theta_1 - \theta_2 \|_2^2 + \| x - \tilde{x} \|_2^2 + \| h_{\theta_1}(x) - h_{\theta_2}(\tilde{x}) \|_2^2) \\
    \leq L_f^2 (1+ L_h^2) (\| \theta_1 - \theta_2 \|_2^2 + \| x - \tilde{x} \|_2^2), \label{lip:2}
\end{multline}
and
\begin{multline} \label{lip:3}
    \| \psi_{\eta_1} (u_{t-n}^{t-1}, y_{t-n}^{t}) - \psi_{\eta_2} (u_{t-n}^{t-1}, y_{t-n}^{t}) \|_2^2  \\\leq L_\psi^2 \| \eta_1 - \eta_2 \|_2^2.
\end{multline}
\end{subequations}
Since $V^{\text{enc},T} (\theta, \eta) = 1/N_\text{sec} \sum_t v_t$ with $N_\text{sec} = N-T-n+1$ as defined in \eqref{eq:encoderloss}, by the sum property, we have that $L_{\text{enc}, T} = L_{v_t}$. Using the relations \eqref{lip:1}-\eqref{lip:3}, it is possible to derive $L_{v_t}$ in terms of $L_h, L_f, L_\psi$ and $T$. A similar derivation has been done by Ribeiro et. al.~\cite{ribeiro2020smoothness} of the cost function $V_T(\theta) \triangleq \frac{1}{T} \sum_{t=1}^T \| y_t - \hat{y}_t \|_2^2$ where an OE noise model was considered and, instead of an encoder, different initial states $x_0$ were used. They showed that the following scaling law applies
\begin{gather}
    L_{V_T} = \mathcal{O}(L_{f}^{2 T});\ \ L_{V_T}' = \mathcal{O}(L_{f}^{3 T}).
\end{gather} 
 when $L_{f}>1$ and where $L_{V_T}'$ represents the Lipschitz constant of the derivative of $V_T$. Hence, to derive the scaling of $L_{v_t}$ and thus $L_{\text{enc}, T}$ we rely on this derivation by only showing that these differences leave the exponential scaling with $T$ unaltered. 
 
 Adapting this result to our considered case is relatively simple to show since the encoder only changes the initial state difference $\| x_0 - \tilde{x}_0 \|_2^2$ to $L_\psi^2 \| \eta_1 - \eta_2 \|_2^2$ which is independent of $T$ and the change to innovation structure replaces $L_{f}$ by $L_{\Tilde{f}} = L_f \sqrt{1+ L_h^2}$. 
  \hfill $\blacksquare$ \vspace{-3mm}
\end{pf} 

\subsection{Data-efficiency with overlapping subsections}
\label{sec:data-efficiency}

To quantify the data-efficiency of overlapping subsections, consider a fixed $T$ for the $T$-step truncated prediction loss and analyze the data efficiency using equidistantly placed sections in terms of the distance parameter $d$, i.e. $\mathcal{I}=\{ 1+d k \in \mathbb{I}_1^{N-T-1} \mid k\in\mathbb{N} \cup 0 \}$. The parameter $d$ regulates the distance between each sub-section where $d=1$ recovers the encoder formulation and $d=T$ recovers the conventional approach in multiple shooting with no overlap. 
To make the notation more compact,  introduce a change of variables in the sum \eqref{eq:l-def} such that
\begin{gather}
    V_{\mathcal{D}_\tx{N}}^{d}(\theta,\eta) = \frac{1}{m_d} \sum_{\tx{k}=0}^{m_d-1} v_\tx{1 + dk}
\end{gather}
where $N-T+1=d m_d +r_d$ with $m_d,r_d\in\mathbb{N}$ and $0\leq r_d<d$. 

To define data-efficiency, we assume stationary input and output signals 

\begin{assum}[Stationarity]\label{assum:y-random}
Both the model output $\hat{y}_{t+k|t}$ and system output $y_t$ are assumed to be strictly statistically stationary. In other words, the cumulative distribution function $p_Y$ of the joint distribution of instances of 
$y_{t}$ at times $t_1, ..., t_n$ has the property that
\begin{equation}
    p_Y(y_{t_1+\tau},...,y_{t_n+\tau}) = p_Y(y_{t_1},...,y_{t_n}).
\end{equation}
for all $t_1, ..., t_n, \tau \in \mathbb{Z}$.
\end{assum}

This assumption is reasonable since many fading memory system like bi-linear systems~\cite{priestley1988bilinear}, and Volterra series~\cite{boyd1985volterra} have the property that the system output $y_t$ is quasi stationary if the input $u_t$ is stationary. However, to our knowledge a proof of this property for stable systems defined via an NL-SS representation is not present in the literature.
Firstly, to show enhanced data efficiency we need that the cost functions converge to the same cost function in the limit of infinite data.

\begin{thm}[Asymp.~insensitivity for $T$] \label{thm:limit-Vd}
Under Assumption \ref{assum:y-random}, both $V_{\mathcal{D}_\tx{N}}^{1}(\theta,\eta)$ and $V_{\mathcal{D}_\tx{N}}^{T}(\theta,\eta)$ converge to the same loss function  with probability 1 when $N \rightarrow \infty$.
\end{thm} \vskip -3mm
\begin{pf}
Assumption \ref{assum:y-random} implies that $v_t$ is also strictly stationary since any signal which is dependent only on strictly stationary variables is also strictly stationary. Furthermore, by the law of large numbers, the infinite mean sum of $v_t$ becomes equal to $\mathds{E} [ v_\tx{k} ]$ with probability 1. Hence, the limit cost can be expressed as \vspace{-1mm}
 \begin{align*}
      \lim_{N \rightarrow \infty} V_{\tx{D}_\tx{N}}^{d}(\theta,\eta) &= \lim_{N \rightarrow \infty} \frac{1}{m_d} \sum_{\tx{k}=0}^{m_d-1} v_\tx{1
    + dk} = \mathds{E} [ v_\tx{k} ]
 \end{align*} 
which is independent of $d$. \hfill $\blacksquare$
\end{pf} 
\vskip -3mm
Next, we show that allowing for overlap, e.g. by taking $d=1$, reduces the variance of the loss function compared to disallowing overlap by $d=T$. 

\begin{thm}[Overlap effect]
With Assumption~\ref{assum:y-random}, there exists an $N_\ast \in\mathbb{N}$, such that, for all $N>N_\ast$, the following relation holds 
$$
\mathrm{Var}(V_{\tx{D}_\tx{N}}^{1}(\theta_\ast,\eta_\ast)) \leq \mathrm{Var}(V_{\tx{D}_\tx{N}}^{T}(\theta_\ast,\eta_\ast))
$$
for all $(\theta_\ast,\eta_\ast) \in \Xi_\ast$ given by Eq. \eqref{eq:xi-ast-def}.
\end{thm}
\begin{pf}
Proving this statement is equivalent to showing that the function 
\begin{gather}
    G(d) \triangleq \text{Var}\!\left (V_{\tx{D}_\tx{N}}^{d}(\theta_\ast,\eta_\ast) \right )\!  = \!\text{Var}\!\left (\frac{1}{m_d}\!\! \sum_{\tx{k}=0}^{m_d-1} v_\tx{1 + dk} \right ) \label{eq:overalp}
\end{gather}
has the property of $G(1) \leq G(T)$ for all $N>N_\ast$. By expanding \eqref{eq:overalp} using conventional variance and covariance relations, we get
\begin{gather}
    G(d) = \frac{1}{m_d^2} \sum_{\tx{k}=0}^{m_d - 1} \sum_{\tx{l}=0}^{m_d - 1} \text{Cov}(v_{1+\tx{k} d},v_{1+\tx{l} d}).
\end{gather}
As we have shown in Theorem \ref{thm:limit-Vd}, $v_t$ is strictly stationary under Assumption \ref{assum:y-random}, hence we can replace the covariance by 
$C(d |\tx{k}-\tx{l}|) \triangleq \text{Cov}(v_{\tx{k} d},v_{\tx{l} d})$ and use the auto-correlation function $R(t) \triangleq C(t)/C(0)$ 
to simplify the expression to 
\begin{gather}\label{eq:Geq}
    G(d) = \frac{1}{m_d^2} \left ( m_d + 2 \sum_{\tx{t}=1}^{m_d - 1} (m_d - t ) R(t d) \right).
\end{gather}

The only part which remains to complete the proof is to determine if $R(\tx{t} d)$ under the given assumptions implies $G(1)\leq G(T)$ using this expression.

The value of $R(\tx{t} d)$ can be derived from 
\begin{gather}
    v_t = \frac{1}{T} \sum_{k=0}^{T-1} \|  y_{t+k} -\hat{y}_\ar{t}{t+k} \|_2^2 = \frac{1}{T} \sum_{k=0}^{T-1} \| e_{t+k} \|_2^2 
\end{gather}
since we evaluate the cost in ($\theta_\ast$,$\eta_\ast$). Hence, 
\begin{gather}
    \text{Cov}(v_t,v_{t+\tau}) \sim \sum_{i=0}^{T-1} \sum_{j=0}^{T-1} \text{Cov}( \| e_{t+i} \|_2^2 , \| e_{t+\tau+j} \|_2^2 )
\end{gather}
where $\text{Cov}( \| e_{t+i} \|_2^2 , \| e_{t+\tau+j} \|_2^2 )$ is nonzero if and only if $i=\tau+j$ and it is the same value for any $t$ since $e_t$ is white. Hence, the auto-correlation function is $R(t) = \max(0,1-t/T)$, which simply counts the number of terms which have the same index in the sum. After substitution of $R(t) = \max(0,1-t/T)$ in \eqref{eq:Geq}, it directly follows that $G(1) \leq G(T)$. \hfill $\blacksquare$
\end{pf}

Under consistency of the estimator, the parameter estimates will converge to ($\theta_\ast$,$\eta_\ast$), hence, this result hold in general. This shows that allowing for overlap in the subsections results in a more efficient estimator.

\section{Simulation study}
\label{sec:numerical}

In this section, we demonstrate the effectiveness of the proposed SUBNET architecture based identification approach in an extensive simulation study. As the method has a number of hyperparameters that can substantially alter its behaviour and performance, hence we investigate the effects of these hyperparameters and also motivate the previously provided  guidelines for choosing them. An evaluation of the subspace encoder method on experimental data is provided in Section~\ref{sec:experimental}.

\subsection{Data-generating system}

The following system  is considered:
\begin{subequations} \label{eq:dat:gen:sys}
\begin{align}
    x_{k+1}^{(1)} &= \frac{x_k^{(1)}}{1.2 + \left (x_k^{(2)} \right )^2} + 0.4 \cdot x_k^{(2)}, \\
    x_{k+1}^{(2)} &= \frac{x_k^{(2)}}{1.2 + \left (x_k^{(1)} \right )^2} + 0.4 \cdot x_k^{(1)} + u_k,               \\
    y_k  &= x_k^{(1)} + e_k,
\end{align}
\end{subequations}
where $x_k = [\begin{array}{cc}x_k^{(1)} & x_k^{(2)} \end{array}]^\top $ denotes the elements of the state vector, $x_0 = 0$ and $e$ is generated by an i.i.d.~white Gaussian noise process resulting in an output SNR of $20$ dB. This noise signal is only present in the training and validation data sets and it is omitted in the test set to accurately measure the performance of the obtained models. The system input $u$ is a white, random, uniformly distributed signal $u_k \sim \mathcal{U}(-2,2)$. $N=10^4$ training, $3\cdot 10^3$ validation and $10^4$ test samples are generated with independent realisations of $e$ and $u$. Note that this system for zero input has two stable equilibria at $x = \pm [0.68\ 0.68]^\top$ and one unstable equilibrium point at $x= [0\ 0]^\top$ (the largest eigenvalue of $\nabla_x f(x,0)|_{x=[0\ 0]^\top}$ is $1.23$). Hence, \eqref{eq:dat:gen:sys} is not strictly contractive, but it is stable in the sense of Assumption \ref{assum:S1} which has been checked numerically.

\subsection{Model estimation}

At this point, the system under study is only disturbed by measurement noise, hence, the SUBNET structure is simplified to an output error noise structure  (i.e. $f_\theta$ does not depend on $\hat{e}_k$). During the analysis, we have varied one hyper-parameter while keeping all the others equal to the following base values: $T = 40$, $n=10$, $n_\mathrm{x} = 4$. All the functions $h_\theta$, $f_\theta$ and $\psi_\eta$ are parametrized using 2 hidden layer feedforward neural networks with $64$ nodes per layer, linear bypass and IO normalization. To compute the model estimate, the Adam optimizer~\cite{kingma2014adam} with a batch size of $256$ has been used with  default learning rate of $10^{-3}$ and early stopping using a validation set. 

We chose a base model order $n_\mathrm{x} = 4$ as it will be shown that \emph{immersion effects} are observed under the considered ANN complexity for $h_\theta$, $f_\theta$ and $\psi_\eta$. Immersion is a well-know phenomenon \cite{Ohtsuka2005,Lee1988}, corresponding to the fact that at the price of introducing extra state variables, the original system can be equivalently represented by less complex state-transition and output functions, till reaching the class of linear, but often infinite dimensional representations, coined in the literature as Koopman forms \cite{williams2016koopman}. As will be shown, this trade-off between $n_\mathrm{x}$ and the layer-depth of the involved ANNs makes the optimisation problem more stable and ensures relatively fast convergence.

The performance of the estimated models is characterised by the \emph{normalized root mean square} (NRMS) simulation error: 
\begin{gather} \label{eq:NRMS}
    \text{NRMS} = \frac{\text{RMS}}{\sigma_\mathrm{y}} = \frac{\sqrt{1/N \sum_{k=1}^{N} \| y_k - \hat{y}_k\|_2^2}}{\sigma_\mathrm{y}},
\end{gather}
where $\sigma_\mathrm{y}$ is the \emph{sample standard deviation} of $y$ and $\text{NRMS}\% = \text{NRMS} \times 100$. In this case, an independent noiseless data set has been used to calculate the NRMS simulation error to clearly characterise the accuracy of the estimated model. The estimated encoder is used to initialize the state for every simulation. Hence, the first $n$ outputs are not used to compute \eqref{eq:NRMS} as they have been used to feed the encoder. 

\subsection{Computational cost and performance}\label{sec:perfcomp}

\begin{figure}
    \centering
    \includegraphics[width=\linewidth]{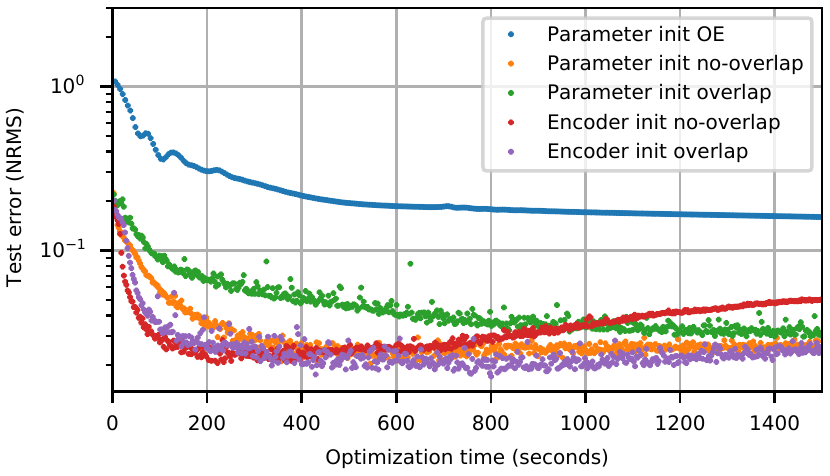}\vspace{-2mm}
    \caption{ Evolution of the NRMS simulation error of the estimated models by the considered approaches w.r.t. the test data. The keywords ``encoder init'' and ``parameter init'' indicate if either encoder-based prediction or parametric estimation is used to estimate the initial states, ``overlap'' and ``no-overlap'' indicate if the subsections can overlap, while ``OE'' stands for simulation based cost over the entire data sequence with no subsections.}
    \label{fig:time-comp}
\end{figure}

The computational cost and performance of the proposed subspace encoder based method is compared to existing methods in \figref{fig:time-comp} and Table \ref{tab:element-compare}. For the comparison, multiple ANN-based state-space identification methods are considered: (i) the classical SS-ANN simulation error minimization method which simulates the entire training range ($T=$ training data length) starting from an initial state which is estimated together with the system parameters (Parameter init OE)~\cite{schoukens2019roadmap}, (ii) the unconstrained multiple-shooting method for SS-ANNs which adds the initial state of each considered simulation section to the parameter vector without (Parameter init no-overlap)~\cite{bock1981multipleshooting} and with (Parameter init overlap) overlapping simulation sections, and finally (iii) the considered subspace encoder method without (Encoder init no-overlap) and with (Encoder init overlap) overlapping simulation sections. 

\figref{fig:time-comp} shows that the "Parameter init OE" approach is indeed the slowest as it needs to simulate the entire training data set to perform one optimization step. It also shows that the encoder provides improved performance compared to parametrization of the initial condition and co-estimating it with the model. Furthermore, the overlap variants of the methods suffer less from overfitting than standard multiple-shooting. This is in line with the variance reduction obtained from the overlapping subsections as shown in Section~\ref{sec:lip}.

\begin{table}[t]
\centering
\caption{Performance of the compared approaches (see Section \ref{sec:perfcomp}) given the same training budget of 25 minutes (and same hardware). \vspace{1mm}}
{\renewcommand{\arraystretch}{1.2}
\begin{tabular}{ll}
\hline
 Combination               & NRMS test   \\
\hline
 Parameter init OE         & 15.9\%       \\
 Parameter init no-overlap & 2.0\%        \\
 Parameter init overlap    & 3.0\%        \\
 Encoder init no-overlap   & 2.1\%        \\
 Encoder init overlap      & 1.7\%        \\
\hline
\end{tabular}
}
\label{tab:element-compare}
\end{table}

\subsection{Truncation length $(T)$}

\begin{figure}
    \centering
    \includegraphics[width=\linewidth]{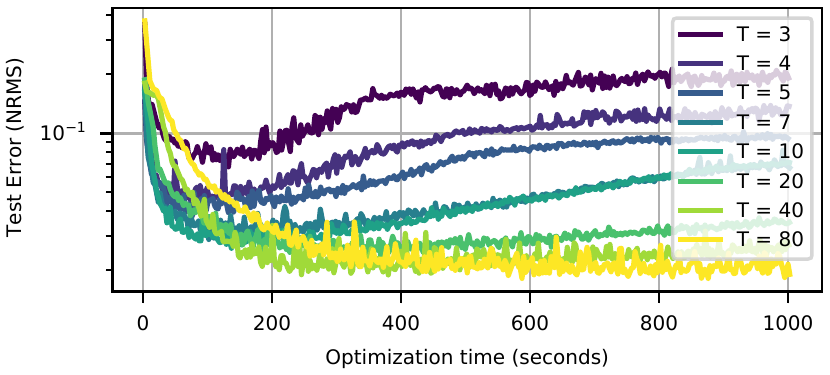} \vspace{-7mm}
    \caption{Influence of the truncation length $T$ of the loss function on the test error during training.}
    \label{fig:nf-influ-time}
\end{figure}

\begin{figure}
    \centering
    \includegraphics[width=\linewidth]{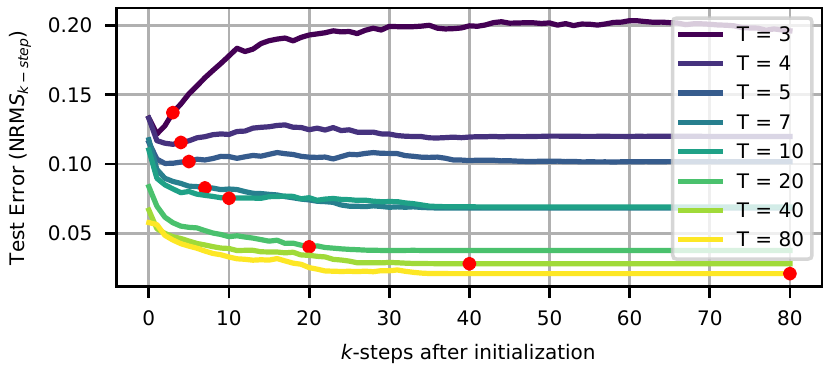} \vspace{-7mm}
    \caption{The $k$-step NRMS error of estimated models under different truncation lengths, computed on the test data. The red dots indicate the truncation length $T$.}
    \label{fig:nf-k-step}
\end{figure}

A key parameter of the encoder method is the truncation length $T$. \figref{fig:nf-influ-time} illustrates that overfitting is a significant issue when $T$ is smaller than the dominant timescale of the system. However, a bigger $T$ increases the computational cost.
To choose $T$ in an informed manner, we employ \emph{expected normalized $k$-step-error} plots in \figref{fig:nf-k-step} in terms of
\begin{equation}
    \text{NRMS}_{\text{k-step}} = \frac{1}{\sigma_\mathrm{y}} \sqrt{\frac{1}{N-k} \sum_{t=1}^{N-k} \| \hat{y}_\ar{t}{t+k} - y_\tx{t+k} \|^2_2 },
\end{equation}
where $\hat{y}_\ar{t}{t+k}$ corresponds to the $k$-step-ahead prediction by the estimated model relative to sample $t$.
 These show that, for the current system under test, the transient error has a length of $30$ time steps, which approximately corresponds to the characteristic timescale of the system. Furthermore, \figref{fig:nf-k-step} shows that the error in the output is larger for low values of $k$ compared to large values of $k$. Hence, the estimated state at $k=0$ is less accurate than the state obtained after a number of steps. This suggests that the encoder is unable to accurately recover/estimate the initial state, i.e.~the encoder estimate has a high variance, indicating that the encoder parametrization is not sufficiently rich ($n = 10$ is insufficient or the used number of layers/neurons is too low).

\subsection{Model order $(n_\mathrm{x})$}

\begin{figure}
    \centering
    \includegraphics[width=\linewidth]{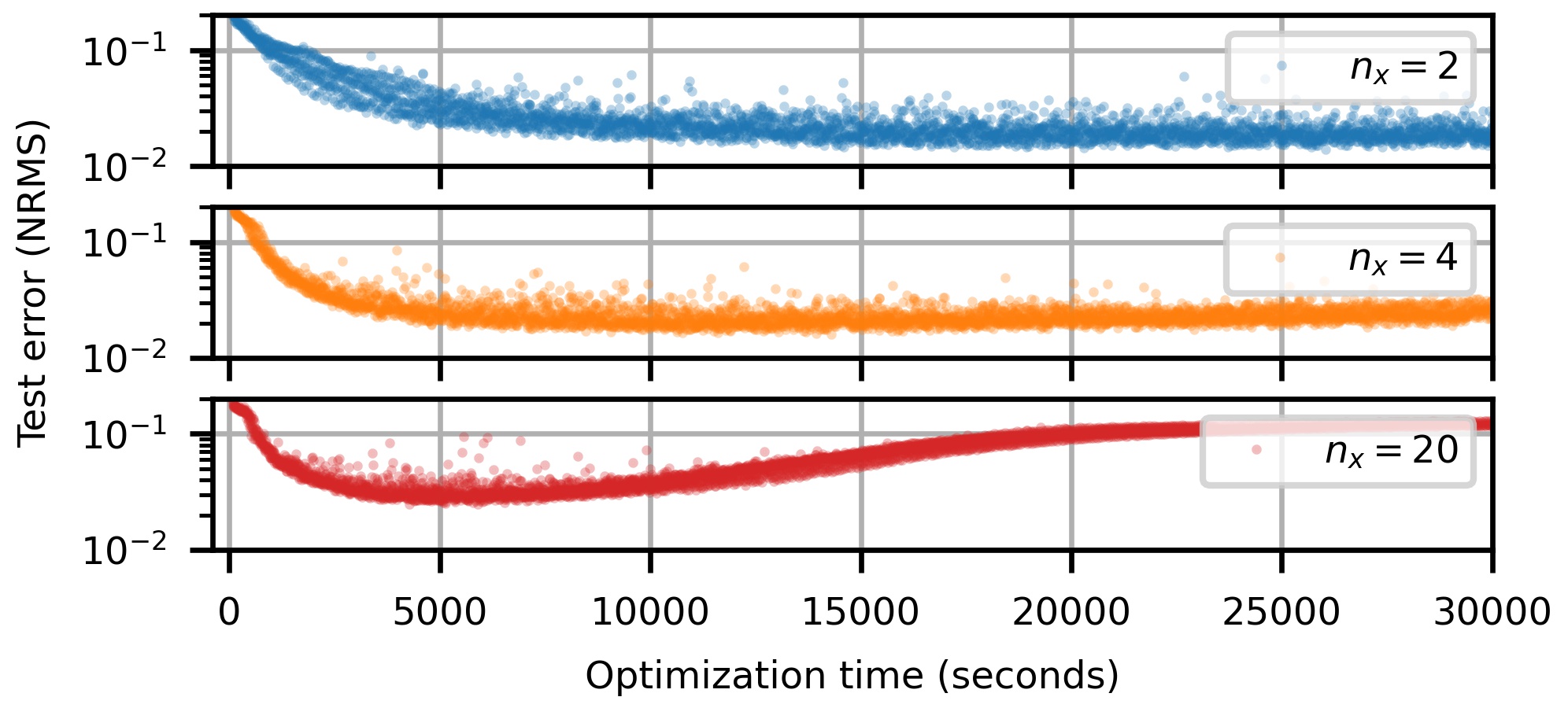} \vspace{-7mm}
    \caption{The influence of the model order $n_\mathrm{x}$ on the resulting test error of the estimates during training. The displayed results are based on 7 models trained using different random initial parameter values.}
    \label{fig:nx-repro}
\end{figure}

Changing the state order of the model has also significant influence on the behaviour of the estimation as shown in \figref{fig:nx-repro}, where the obtained results are averaged over 7 identification runs. $n_\mathrm{x}$ values that are much larger than the order of the true data-generating system quickly result in overfitted model estimates. Using the true state dimension $n_\mathrm{x}=2$ results in a model with the lowest obtained NRMS, but the variability of the obtained models over the 7 identification runs is quite high, and the optimization time is significantly larger than for $n_\mathrm{x}=4$. As discussed earlier, we suspect that this can be attributed to an effect similar to immersion where providing additional degrees of freedom makes the functions $f_\theta$, $h_\theta$ and $\psi_\eta$ less complex and hence simpler to estimate up to the point where the variance of the sheer number of extra parameters overtakes this advantage. 

The 7 different runs, each initialised with different random parameters, converge to the same test error level in NRMS. This suggest that the encoder method gives reproducible results for different initializations which highlights insensitivity of the method for initialisation. 

\subsection{Encoder window length $(n)$}

\begin{figure}
    \centering
    \includegraphics[width=\linewidth]{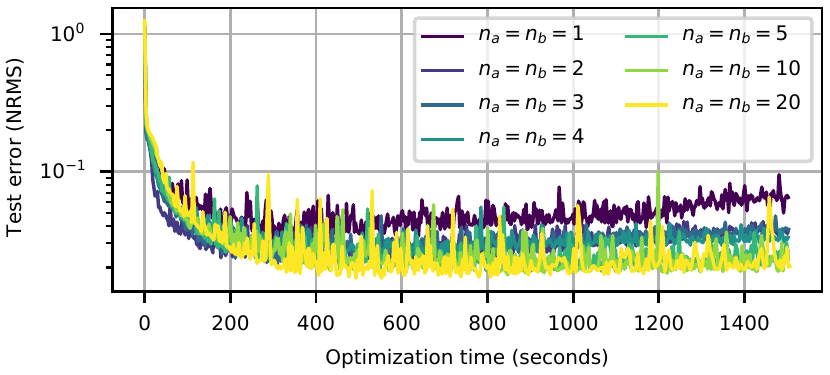} \vspace{-7mm}
    \caption{The influence of the lag window $n_\mathrm{a}$ and $n_\mathrm{b}$, i.e. the horizon of past data used by %
    the encoder, on the test error during training.}
    \label{fig:nanb-fig}
\end{figure}

Varying the lag window $n$ (or $n_\mathrm{a}$ and $n_\mathrm{b}$ considered separately for $u$ and $y$) of the encoder in terms of the considered past IO samples shows in \figref{fig:nanb-fig} that the minimal required $n=n_\mathrm{a}=n_\mathrm{b}=n_\mathrm{x}-1=1$, with $n_\mathrm{x}$ the order of the system for state reconstrutability, on which the encoder is based on, is not the optimal value in this case. As seen from the figure, $20 \geq n=n_\mathrm{a}=n_\mathrm{b}>4$ perform  much better. As we argued earlier, this can be contributed to a variance reduction in the state estimate (i.e. encoder as a minimum variance observer) which reduces the average transient error. However, for values that are much larger than $n_\mathrm{x}-1=1$ (around $20$), the computational cost significantly increases without too much gain in performance. Lastly, the choice of lag windows has a less strong impact on performance than the choice of $T$ and $n_\mathrm{x}$. 

\subsection{Neural network depth and width}

\begin{figure}
    \centering
    \includegraphics[width=0.75\linewidth]{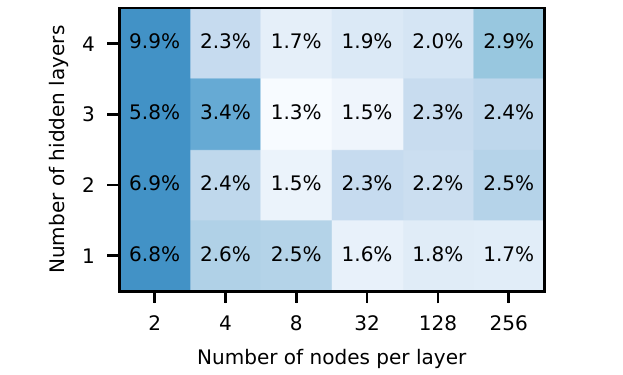} \vspace{-3mm}
    \caption{The influence of the neural network architecture on the model estimates in terms of the achieved NRMS simulation error on test data.}
    \label{fig:struc}
\end{figure}

The influence of varying the depth and width of the neural networks used to parametrize $\psi_\eta$, $f_\theta$ and $h_\theta$ is illustrated in \figref{fig:struc}. A network structure that is too complex results in a growing computational cost and variance of the model estimates. Whereas a network structure that is too simple results in under-fitting as the model is unable to capture the dynamics of the true underlying system. Overfitting is suppressed in our approach by the regularization introduced with early stopping, the stochastic gradient descent algorithm, and the increased data efficiency allowing for overlap in the subsections. Hence, neural network structure selection is more sensitive to underparametrization than overparametrization for the proposed method.

\subsection{Estimation under process noise}

The subspace-encoder based estimation approach has been introduced to provide reliable model estimates under general noise conditions that can be described in an innovation form. To demonstrate this property, we extend \eqref{eq:dat:gen:sys} with a process noise term.
 First we will consider the case when the noise enters the state equation linearly, then we will consider the case when the noise enters nonlinearly:
\begin{subequations}
\begin{align}
    x_{k+1}^{(1)} &= \tfrac{x_k^{(1)}}{1.2 + \left (x_k^{(2)} \right )^2} \!+\! 0.4  x_k^{(2)} \!+\! g_1(x_{k}^{(1)}, e_k),\\
    x_{k+1}^{(2)} &= \tfrac{x_k^{(2)}}{1.2 + \left (x_k^{(1)} \right )^2} \!+\! 0.4  x_k^{(1)} \!+\! u_k +g_2(x_{k}^{(2)}, e_k), \\
    y_k &= x_k^{(1)} + e_k,
\end{align}
\end{subequations}
where $g_i(x_{k}^{(i)},e_k) = K^{(i)}e_k$ for the linear and $g_i(x_{k}^{(i)},e_k) = K^{(i)} x_{k}^{(i)}e_k$ for the nonlinear case respectively, $K \triangleq \sigma_\mathrm{K} K_0/\|K_0\|_2 $ with $K_0 \triangleq [\begin{array}{cc} 1.0 & -0.9 \end{array}]^\top$, and $\sigma_\mathrm{K} \geq 0$ is an adjustable parameter which regulates how strong the process noise is affecting the state. The noise $e_k$ is generated by a white Gaussian noise process with a standard deviation $\sigma_\mathrm{e} = 0.082$ (resulting in 20dB SNR) for the training data and $\sigma_\mathrm{e}=0$ for the test data.

We compare estimation with the subspace encoder method under three different noise models:  \textit{OE noise:} there is no process noise considered in the model, \textit{linear innovation:}   $\hat{e}_{t+k|t}$ appears linearly in $f_\theta$, i.e. in the state equation \eqref{eq:model:state} of the model and \textit{general innovation:}  $\hat{e}_\tx{t+k|t}$ is passed through the neural network $f_\theta$. The resulting NRMS of the simulation error over the test data for the three considered models can be viewed in Table~\ref{tab:linear-inno} and Table~\ref{tab:nonlinear-inno} under linear and nonlinear innovation noise in the data-generating system, respectively. To indicate the significance of the results these tables also contain the standard deviation of the mean which is the sample standard deviation divided by $\sqrt{4-1}$ since there are 4 independent samples considered. The linear case shows that for $\sigma_\mathrm{K}>0.5,$ both the linear and the nonlinear parametrization of the process noise model structure significantly reduces the test error in comparison with the OE model. For the nonlinear case, the nonlinear parametrization outperforms the linear innovation noise model structure for $\sigma_\mathrm{K}>0.5$, as expected. 

\begin{table}[t]
\centering
\caption{Mean and mean standard deviation of the NRMS\% of the simulation error on the test data over 4 independent runs, when \textbf{linear} innovation noise is present in the data-generating system.} \vspace{1mm}
{\renewcommand{\arraystretch}{1.0}
\begin{tabular}{l|lll}
\hline
 $\sigma_\mathrm{K}$ & OE noise & \tripleline{Linear \\ innovation \\ noise model} & \tripleline{Nonlinear \\ innovation \\ noise model}  \\ \hline
 0.0     & 1.8$\pm$0.1  & \textbf{1.7}$\pm$0.1 & 1.8$\pm$0.1 \\
 0.25    & 2.1$\pm$0.2  & \textbf{1.9}$\pm$0.1 & 2.1$\pm$0.1 \\
 0.5     & 2.3$\pm$0.1  & \textbf{1.9}$\pm$0.1 & 2.1$\pm$0.1 \\
 1.0     & 3.2$\pm$0.1  & 2.2$\pm$0.1 & \textbf{2.0}$\pm$0.2 \\
 2.0     & 4.8$\pm$0.1  & 3.0$\pm$0.1 & \textbf{2.5}$\pm$0.0 \\
 4.0     & 8.4$\pm$0.3  & 5.7$\pm$0.4 & \textbf{4.4}$\pm$0.1 \\
\hline
\end{tabular}
}
\label{tab:linear-inno}
\end{table}
\begin{table}[t]
\centering
\caption{Same as Table \ref{tab:linear-inno} but for \textbf{nonlinear} innovation noise.} \vspace{1mm}
{\renewcommand{\arraystretch}{1.0}
\begin{tabular}{l|lll}
\hline
 $\sigma_k$ & OE noise & \tripleline{Linear \\ innovation \\ noise model} & \tripleline{Nonlinear \\ innovation \\ noise model}  \\ \hline
0.0     & 1.8$\pm$0.1  & \textbf{1.7}$\pm$0.0  & 1.8$\pm$0.1  \\
0.25    & 2.2$\pm$0.1  & 2.2$\pm$0.1  & \textbf{1.8}$\pm$0.1  \\
0.5     & 2.8$\pm$0.1  & 2.4$\pm$0.1  & \textbf{1.8}$\pm$0.1  \\
1.0     & 4.1$\pm$0.1  & 3.2$\pm$0.1  & \textbf{2.3}$\pm$0.2  \\
2.0     & 8.2$\pm$0.3  & 6.5$\pm$0.4  & \textbf{4.3}$\pm$0.3  \\
4.0     & 14.2$\pm$0.8 & 12.6$\pm$0.6 & \textbf{10.5}$\pm$0.3 \\
\hline
\end{tabular}
}
\label{tab:nonlinear-inno}
\end{table}

\section{Benchmark results}
\label{sec:experimental}

The Wiener--Hammerstein benchmark~\cite{schoukens2009WHbenchmark} is an electronic circuit with a diode-resistor nonlinearity and has been used in benchmarking a wide variety of nonlinear system identification methods. As in~\cite{beintema2021base-encoder}, we split the data set into 80,000 training, 20,000 validation and 78,000 test samples. Similarly as before, we utilize the same network structure and OE noise structure for all three networks in the model, but now with a single hidden layer with 15 hidden nodes and a linear bypass. The hyperparameters used are $n_\mathrm{x} = 6$, $T=80$, $n=n_\mathrm{a} = n_\mathrm{b} = 50$, a batch size of $1024$, early stopping, input/output normalization, and the Adam optimizer with default learning rate of $\alpha = 10^{-3}$.

The obtained results are displayed in Table \ref{tab:result} together with results achieved by other approaches on this benchmark. Many of these methods have been developed for Wiener--Hammerstein identification problems and use varying level of structural knowledge about the system. Remarkably, the proposed approach is able to get the lowest test error results using no structural knowledge or guided initialisation as some of the methods that are reported on this benchmark. Note however that, while good performance is obtained quite rapidly, a long optimization (over 200 hours) is required to fine tune the estimate and obtain the final result. This aspect can be further improved by using learning rate schedulers and higher-order optimization methods.

\begin{table}[t]
\centering
\caption{Results of the subspace encoder on the Wiener--Hammerstein benchmark compared to results reported in the literature.} \vspace{1mm}
{\renewcommand{\arraystretch}{1.2}
\begin{tabular}{l|ll}
\hline
Identification Method               & \tripleline{Test RMS\\ Simulation \\ ($\si{mV}$)} & \begin{tabular}[c]{@{}l@{}}Test NRMS\\ Simulation\end{tabular} \\ \hline
\textbf{Subspace Encoder}      & \textbf{0.241}                                                     & \textbf{0.0987\%}                                               \\
QBLA \cite{app:QBLA-schoukens2014identification} & 0.279                                                           & 0.113\%                                                        \\
Pole-zero splitting \cite{app:pole-splitting-sjoberg2012identification}  & 0.30                                                               & 0.123\%                                                        \\
NL-LFR \cite{schoukens2020NN-state-space-LRF} & 0.30                                                               & 0.123\%                                                        \\
PNLSS \cite{app:PNLSS-marconato2009identification} & 0.42                                                               & 0.172\%                                                        \\
{Generalized WH \cite{app:gen-WH-wills}}      & 0.49                                                               & 0.200\%                                                        \\
LS-SVM \cite{app:ls-svm-falck2009identification}     & 4.07                                                               & 1.663\%                                                        \\
{Bio-social evolution  \cite{app:bio-naitali2016wiener}} & 8.55                                                               & 3.494\%                                                        \\
{SS auto-encoder \cite{masti2021autoencoders}} & 12.01                                                              & 4.907\%                                                        \\
{Genetic Programming \cite{app:GP-Dhruv}} & 23.50                                                              & 9.605\%                                                        \\
{SVM \cite{app:svm-marconato2009identification}}                 & 47.40                                                              & 19.373\%                                                       \\
BLA \cite{app:lin-lauwers2009modelling}                  & 56.20                                                              & 22.969\%                                                      
\end{tabular}
}
\label{tab:result}
\end{table}

\section{Conclusion}
\label{sec:conclusion}

By carefully combining elements from machine learning (batch optimization), multiple-shooting (multi-step-ahead loss), and subspace identification (encoder), a subspace encoder-based ANN method for nonlinear state-space identification has been introduced. The method is proven to be locally consistent, to have a relatively smooth cost function by using a multiple shooting strategy, and to be more data efficient than other ANN-based strategies by considering overlapping subsections. The method has shown state-of-the-art results in our simulation studies and on a widely used benchmark identification problem. Remarkably, it does not need specialized initialization of the model parameters to achieve this performance or structural knowledge of the system and besides of some rules of thumb, it is also relatively insensitive to the choices of its hyper parameters. 

\bibliographystyle{ieeetr}        %

\bibliography{references}  

\end{document}